\documentclass[twocolumn,aps,superscriptaddress,showpacs,floatfix,prd,noshowpacs]{revtex4-2}
\usepackage[body={19.cm,24.cm}]{geometry}     
\usepackage{mathrsfs}
\usepackage{amssymb}
\usepackage{amsmath}
\usepackage{graphicx}
\usepackage[normalem]{ulem}
\usepackage[dvips]{color}
\usepackage{bm}
\usepackage{longtable}
\usepackage{slashed}
\usepackage{enumitem}
\usepackage{empheq}

\usepackage{titlesec}
\renewcommand{\thesection}{\Roman{section}}
\titleformat{\section}{\small\bfseries\centering}{\thesection.}{0.5em}{}

\usepackage[T1]{fontenc}
\usepackage{utopia}
\usepackage[utopia]{mathdesign}

\usepackage[breaklinks=true]{hyperref}
\hypersetup{
  colorlinks=true,
  citecolor=magenta,
  linkcolor=black,
  urlcolor=teal,
}

\renewcommand\sout{\bgroup \color{red} \ULdepth=-.5ex \ULset}

\renewcommand{\rm}[1]{\textrm{#1}}
\renewcommand{\d}{\mathrm{d}}

\usepackage{tikz,xcolor,hyperref}

\definecolor{lime}{HTML}{A6CE39}
\DeclareRobustCommand{\orcidicon}{
	\begin{tikzpicture}
	\draw[lime, fill=lime] (0,0) 
	circle [radius=0.16] 
	node[white] {{\fontfamily{qag}\selectfont \tiny ID}};
	\draw[white, fill=white] (-0.0625,0.095) 
	circle [radius=0.007];
	\end{tikzpicture}
	\hspace{-2mm}
}
\foreach \x in {A, ..., Z}{%
	\expandafter\xdef\csname orcid\x\endcsname{\noexpand\href{https://orcid.org/\csname orcidauthor\x\endcsname}{\noexpand\orcidicon}}
}

\begin{document}

\title{An Effective Upper Bound on the Pressure-to-Energy Density Ratio in Neutron Stars}


\author{Bao-Jun Cai\orcidA{}}\email{bjcai@fudan.edu.cn}
\affiliation{Key Laboratory of Nuclear Physics and Ion-beam Application (MOE), Institute of Modern Physics, Fudan University, Shanghai 200433, China} 
\affiliation{Shanghai Research Center for Theoretical Nuclear Physics, NSFC and Fudan University, Shanghai 200438, China}
\author{Bao-An Li\orcidB{}}\email{Bao-An.Li$@$etamu.edu}
\affiliation{Department of Physics and Astronomy, East Texas A\&M University, Commerce, TX 75429-3011, USA}
\author{Yu-Gang Ma\orcidC{}}\email{mayugang$@$fudan.edu.cn}
\affiliation{Key Laboratory of Nuclear Physics and Ion-beam Application (MOE), Institute of Modern Physics, Fudan University, Shanghai 200433, China} 
\affiliation{Shanghai Research Center for Theoretical Nuclear Physics, NSFC and Fudan University, Shanghai 200438, China}
\affiliation{College of Physics, East China Normal University, Shanghai 200241, China}

\date{\today}

\newcommand{\x}{\mathrm{X}}
\newcommand{\y}{\mathrm{Y}}
\newcommand{\hr}{\widehat{r}}
\newcommand{\hP}{\widehat{P}}
\newcommand{\heps}{\widehat{\varepsilon}}
\newcommand{\hrho}{\widehat{\rho}}

\fontdimen2\font=1.5pt

\begin{abstract}

The equation-of-state (EOS) parameter $\phi \equiv P/\varepsilon$, defined as the ratio of pressure to energy density, encapsulates the fundamental response of matter under extreme compression. Its value at the center of the most massive neutron star (NS), $\x \equiv P_{\rm c}/\varepsilon_{\rm c}$, provides an upper bound on the maximum attainable central EOS parameter of cold visible matter. Remarkably, owing to the intrinsically nonlinear structure of the EOS in General Relativity (GR), this bound lies far below the naive Special Relativity (SR) limit of unity. In this work, we refine the theoretical upper bound on $\x$ in a self-consistent manner by incorporating, in addition to the causality constraint from SR, the mass-sphere stability condition associated with the mass evolution pattern in the vicinity of the NS center. This condition is formulated within the intrinsic and perturbative analysis of the dimensionless Tolman--Oppenheimer--Volkoff equations (IPAD-TOV) framework. The combined constraints yield an improved bound, $\x \lesssim 0.385$, which is slightly above but fully consistent with the previously derived causal-only limit, $\x \lesssim 0.374$. We further derive an improved scaling relation for NS compactness and demonstrate its robustness across a broad set of 284 realistic EOSs, including models with first-order phase transitions, exotic degrees of freedom, continuous crossover behavior, and deconfined quark cores. Within the IPAD-TOV framework, the resulting bound on $\x$ provides a new EOS-insensitive probe of the microphysics of cold superdense matter compressed by strong-field gravity in GR.

\end{abstract}

\maketitle

\section{Introduction}

Neutron stars (NSs) host the densest visible matter in our Universe, which provide a unique laboratory to probe strongly interacting matter under extreme densities and strong-field gravity in General Relativity (GR). The cold dense matter equation of state (EOS), $P = P(\varepsilon)$, which relates the pressure $P$ to the energy density $\varepsilon$, governs the internal structure and global properties of NSs\,\cite{Shapiro1983}, determining key observables such as the mass-radius (M-R) relation, tidal deformability, and the maximum mass. 
The EOS also plays a central role in interpreting heavy-ion collision experiments\,\cite{LCK08,Lovato22,Soren23,WeiSN24,ZhaoJ24,Shou24,JHChen25,LiuYY25}, nuclear structure studies\,\cite{Bend03,Stone07,Ding24,WeiK24,XuY24,Qin25,AnR24}, QCD-based investigations of dense matter\,\cite{Alford2008,Baym18}, and the composition of dense matter including hyperonic degrees of freedom\,\cite{Isa18,Tolos20,Sed23}, as well as astrophysical observations of NSs through supernovae\,\cite{Oertel17,Wat16}, NS mergers\,\cite{Bai19,Chat24,Alar25,LiA25}, gravitational waves\,\cite{Abbott2017,Abbott2018,Abbott2020-a,Bai17}, and X-ray measurements\,\cite{Riley19,Miller19,Fon21,Riley21,Miller21,Salmi22,Choud24,Reardon24,Mauv25}.
Despite decades of intensive work on the EOS $P(\varepsilon)$\,\cite{Tews18,Baym19,McL19,Zhao20,Tan22,Tan22-a,Alt22,Dri22,Huang22,Ecker22,Ecker23,Pro23,Som23,
Ann18,Ann23,Ess21,Brandes23,Brandes23-a,Tak23,Pang23,Fan23,Bed15,Olii23,ZLi25,Komo23,Marc23,Jim24,LBA24,ZNB25,Grun25-a,LBA25,Ecker25,Huang25,Patra25,Ng25,Bisw25,Reed25,Pass25,Wout25,Baus25,Tang25,ZLi23,Cao23,Mro23,Semp25,Cuce25,Ferr24,Ferr25,Fuji24,Wang24,Ye25,CuiY25,Som23-a,Zhou25,Marc24,Zhou19,Jiang20,Shao20,Raai21,Tang21,Muso24,Koehn24,Snep24,Saes24,Ann22,LCXZ21,Latt21,Dri21} and the wealth of observational constraints especially since GW170817, surprisingly little attention has been paid to the dimensionless EOS parameter $\phi \equiv P/\varepsilon$, which uniquely determines the trace anomaly through $\Delta = 1/3 - \phi$\,\cite{Fuji22}. This quantity characterizes NS compactness\,\cite{CL24-a,CL24-b,CL25-b,CLM25-phi} and represents the energy-density--averaged squared speed of sound up to the energy density $\varepsilon$\,\cite{Marc24,Saes24}. Investigating $\phi$ therefore provides new insight into the internal structure of NSs beyond the conventional EOS relation $P(\varepsilon)$\,\cite{Mous1,Mous2,Li26}. Moreover, it enables the evolution of the trace anomaly with energy density to be tracked in both NSs and heavy-ion collisions within an approximately universal framework\,\cite{Li26}. Interestingly, while the EOS parameter $\phi$ has long been studied in cosmology to understand the expansion dynamics driven by radiation, matter, and dark energy\,\cite{Carr01,Peeb03,Cope06,Koma11,Aubo15,Agha20}, its role in NSs and heavy-ion collisions has only recently begun to receive focused attention; see, for example, Refs.\,\cite{Fuji22,CL24-a,CL24-b,CL25-b,CLM25-phi,Mous1,Mous2,Li26}.

To study the ratio $\phi$ in a nearly EOS-model-independent manner, we apply the IPAD-TOV approach\,\cite{CLZ23-a,CLZ23-b,CL25-a} which conducts an Intrinsic-and-Perturbative Analysis of the Dimensionless Tolman--Oppenheimer--Volkoff (TOV) equations\,\cite{TOV39-1,TOV39-2,Misner1973}. In GR, the TOV equations describe the radial balance of nuclear pressure and gravity in static, spherically symmetric NSs under general relativistic hydrodynamic equilibrium. 
By introducing reduced variables normalized to the central energy density, the TOV equations can be expressed in a dimensionless form, revealing intrinsic relations among the coefficients of polynomial expansions of the pressure and energy density in terms of the reduced radial coordinate. These relations enable the extraction of central EOS information directly from NS observables such as mass and radius, without invoking a specific input EOS model. Although the underlying perturbative analysis is formulated locally near the stellar center, the resulting scaling relations have been found to remain accurate when compared with full TOV solutions using a broad range of realistic EOSs\,\cite{CLZ23-a,CLZ23-b,CL24-a,CL24-b,CL25-b,CL25-a}. They have subsequently provided new insights into NS core properties, e.g., the causality boundary of the M-R curve\,\cite{CLZ23-b}.
Recently, the IPAD-TOV method has been applied to study the behavior of the EOS-parameter $\phi$ near the NS center\,\cite{CLM25-phi}, showing that $\phi$ reaches its maximum at the center, i.e., $\phi \le \x \equiv {P_{\rm c}}/{\varepsilon_{\rm c}}$. Consequently, the central value $
\x$
provides an effective upper bound for the $\phi$ attainable in any visible matter in the Universe. Using the IPAD-TOV framework, we previously found that $\x \lesssim 0.374$ by imposing the causality condition of Special Relativity (SR)\,\cite{CLZ23-a}, namely requiring the sound-speed squared never exceeds unity.

In this work, we extend the IPAD-TOV analysis to refine the upper bound on the central EOS-parameter $\x$ by incorporating, in addition to the causality limit, the mass-sphere stability condition associated with the mass evolution pattern near the NS center. Using this framework, we obtain a refined bound about $\x \lesssim 0.385$, slightly higher but consistent with the previous causality-only constraint $\x \lesssim 0.374$, and demonstrate its implications for improved NS compactness scaling relations across a broad set of EOS models.
The upper bound on $\x$ provides a characteristic scale for dense matter in NS interiors. Determining this limit with high precision may offer new insights into the properties of superdense matter. Moreover, further refining its value may establish a useful benchmark for dense-matter physics under strong gravitational fields and improve our understanding of matter at the highest densities, where GR, quantum many-body physics, and Quantum Chromodynamics collectively govern NS structure\,\cite{Fuji22,Bjorken83,Kurkela10,Gorda21PRL,Gorda23PRL,Gorda23,Komo22,Braun2022,Brandt2023,Fuku2024}.

The rest of this paper is organized as follows. In Section \ref{SEC_IPADTOV}, we briefly review the IPAD-TOV approach, highlighting the aspects most relevant to this work. In Section \ref{SEC_a2}, we introduce a Gedankenexperiment and analyze the physical information contained in the first nontrivial expansion of the NS energy density within the IPAD-TOV framework, revealing the mass-sphere stability condition in addition to the causality requirement from SR. Section \ref{SEC_EFFC} then evaluates the effective correction to the upper bound of the central EOS-parameter $\x$, while Section \ref{SEC_AT} explores alternative forms of this correction, complementing the analysis of the previous section. Finally, we summarize our main findings in Section \ref{SEC_SUM}.

\section{Brief Review of the IPAD-TOV Method}\label{SEC_IPADTOV}

In this section, we outline the IPAD-TOV approach\,\cite{CLZ23-a,CLZ23-b,CL25-a}, 
highlighting the features most relevant for investigating the central EOS-parameter $\x$.  
In units $c=G=1$, the dimensionless TOV equations read\,\cite{CLZ23-a}
\begin{equation}
\frac{\d\widehat{P}}{\d\widehat{r}} 
= - \frac{\widehat{\varepsilon}\widehat{M}}{\widehat{r}^2} 
\frac{(1+\widehat{P}/\widehat{\varepsilon}) (1 + \widehat{r}^3 \widehat{P}/\widehat{M})}{1 - 2\widehat{M}/\widehat{r}}, \quad
\frac{\d\widehat{M}}{\d\widehat{r}} = \widehat{r}^2 \widehat{\varepsilon},
\end{equation}
where the reduced variables are defined as 
$\widehat{P}=P/\varepsilon_{\rm c}$, $\widehat{\varepsilon}=\varepsilon/\varepsilon_{\rm c}$,
$\widehat{r}=r/Q$, and $\widehat{M}=M/Q$.  
The characteristic length and mass scale $Q$ is defined as\,\cite{CL25-a}
\begin{equation}
Q=\frac{1}{\sqrt{4\pi\varepsilon_{\rm c}}}
\approx 10\cdot\left(\frac{\varepsilon_{\rm c}\text{ in MeV/fm}^3}{600}\right)^{-1/2}\,\rm{km},
\end{equation}
so that $Q$ is generically of order $\mathcal{O}(10\,\rm{km})$.
The reduced radius $\widehat{R}$ is determined by $\widehat{P}(\widehat{R})=0$ on NS surface, and the NS mass follows as 
\begin{equation}
\widehat{M}_{\rm{NS}}=\widehat{M}(\widehat{R})=\int_0^{\widehat{R}} \d\widehat{r}\,\widehat{r}^2\widehat{\varepsilon},
~\text{or equivalently }M_{\rm{NS}}=M(R).
\end{equation}

Near the center, two small quantities naturally arise: the reduced radius $\widehat{r}$ (or $\mu\equiv\widehat{\varepsilon}-1$) 
and the central EOS-parameter $\x=\widehat{P}_{\rm c}<1$\,\cite{CLM25-phi}.  
These allow a general double-element expansion of the relevant stellar quantity $\mathcal{U}$\,\cite{CL25-a}:
\begin{equation}
\mathcal{U}/\mathcal{U}_{\rm c}
\approx 1+\sum_{i+j\ge1}u_{ij}\x^i\widehat{r}^j,
\end{equation}
where $\mathcal{U}_{\rm c}$ is the $\mathcal{U}$ at the center, and the coefficients $\{u_{ij}\}$ follow from the TOV equations.  
Low-order coefficients are universal, independent of the input EOS, which provides a model-insensitive description of the NS core\,\cite{CLZ23-b}.  
In the limit $\widehat{r}\to0$, the expansion becomes exact.

For the pressure, the perturbative expansion is\,\cite{CLZ23-a}
\begin{equation}
\widehat{P}(\widehat{r})\approx\x+b_2\widehat{r}^2+b_4\widehat{r}^4+\cdots,
\end{equation}
where\,\cite{CLZ23-a}
\begin{equation}
b_2=-\frac{1}{6}(1+3\x^2+4\x),~~
b_4=-\frac{1}{2}b_2\!\left(\x+\frac{4+9\x}{15s_{\rm c}^2}\right),
\end{equation}
and $s_{\rm c}^2=\d\widehat{P}/\d\widehat{\varepsilon}|_{\rm c}=b_2/a_2$ is the central speed of sound squared.  
The energy-density expansion is similarly
\begin{equation}
\widehat{\varepsilon}(\widehat{r})\approx1+a_2\widehat{r}^2+a_4\widehat{r}^4+\cdots.
\end{equation}
By symmetry, only even powers appear in expanding $\widehat{P}(\widehat{r})$ and $\widehat{\varepsilon}(\widehat{r})$\,\cite{CLZ23-a,CLZ23-b}.  
The coefficients satisfy $b_2<0$, $b_4>0$, and $a_2=b_2/s_{\rm c}^2<0$, while $a_4$ may take either sign.  
All coefficients are naturally $\mathcal{O}(1)$.

Keeping only $\mathcal{O}(\widehat{r}^2)$ terms gives the reduced radius scaling\,\cite{CLZ23-a}
\begin{equation}
\widehat{R}^2\sim\frac{\x}{1+3\x^2+4\x}.
\end{equation}
For consistency with our previous works, we denote the above quantity by $\Pi_{\rm c}$\,\cite{CL25-b}. The corresponding NS radius and mass scalings are then given by\,\cite{CL25-b}:
\begin{align}
R\sim&\frac{\Pi_{\rm c}^{1/2}}{\sqrt{\varepsilon_{\rm c}}},~~
M_{\rm{NS}}\sim\frac{\Pi_{\rm c}^{3/2}}{\sqrt{\varepsilon_{\rm c}}}.
\end{align}
Consequently, the NS compactness scales as $\xi\equiv M_{\rm{NS}}/R\sim\Pi_{\rm c}$.
These relations link global NS observables directly to the central EOS-parameter $\x$, with no dependence on higher-order EOS coefficients such as $a_4$\,\cite{CLZ23-a}.  
They have been validated using hundreds of microscopic and phenomenological EOSs available in the literature, as well as $10^5$ meta-model EOSs\,\cite{Lat24-talk,CL25-b}, confirming their robustness.

By introducing the log-stability slope\,\cite{CLM25-phi}
\begin{equation}
\Psi = 2\frac{\d\ln M_{\rm{NS}}}{\d\ln\varepsilon_{\rm c}},
\end{equation} 
and using the scaling of $M_{\rm{NS}}$, one obtains the $s_{\rm c}^2$ in normal NSs of mass M$_{\rm{NS}}$ as\,\cite{CL24-a}:
\begin{equation}\label{sc2-gen}
s_{\rm c}^2=\x\left(1+\frac{1+\Psi}{3}
\frac{1+3\x^2+4\x}{1-3\x^2}\right).
\end{equation}
Stable configurations satisfy $\Psi\ge0$, and the maximum-mass configuration (where the NS mass peaks on a given mass-radius sequence) corresponds to $\Psi=0$, or $
{\d M_{\rm{NS}}}/{\d\varepsilon_{\rm c}}=0$.
Requiring $s_{\rm c}^2\le1$ at the maximum-mass point yields $\x\lesssim0.374$.  
This reflects the strong nonlinear behavior of $s_{\rm c}^2$ at high density, which significantly lowers the physical limit on the EOS-parameter $\phi=P/\varepsilon$ compared with the naive causal bound from SR.
Since our purpose is to refine the upper bound on $\x$, we focus on the maximum-mass configurations with $\Psi=0$.

For the ease of our later discussion, we rewrite below several relations.  
Using the general NS mass scaling relation,
\begin{equation}
M_{\rm{NS}}\sim
\frac{\widehat{R}^3}{\sqrt{\varepsilon_{\rm c}}}
\sim \frac{1}{\sqrt{\varepsilon_{\rm c}}}\left(\frac{\x}{B(\x)}\right)^{3/2},~~B(\x)\equiv -b_2(\x),
\end{equation}
we can rewrite the $s_{\rm c}^2$ for TOV NSs:
\begin{align}\label{def-sc2BB}
s_{\rm c}^2(\x)
=&\left.\frac{4\x}{3}\left(1-\frac{3}{4}\frac{\d\ln B}{\d\ln\x}\right)\right/
\left(1-\frac{\d\ln B}{\d\ln\x}\right)
=\x\left(1-\frac{1}{3}\frac{\d\ln\x}{\d\ln R}\right).
\end{align}
This expression is quite general. For example, in the case of Newtonian stars, $B$ does not depend on $\x$, and the stellar structure can be analyzed using the Lane--Emden equation (with the Newtonian coefficient $B_{\rm N}=1/6$)\,\cite{Shapiro1983}; consequently, $s_{\rm c}^2\approx4\x/3$.
For small $\x$, one finds
\begin{equation}
s_{\rm{c}}^2
\approx\frac{4\x}{3}\left[1+\frac{1}{4}\left(\frac{\d \ln B}{\d\ln\x}\right)+\frac{1}{4}\left(\frac{\d \ln B}{\d\ln\x}\right)^2+\cdots\right].
\end{equation}
Expanding the coefficient as $B\approx B_{\rm{N}}(1+k_1\x+k_2\x^2+\cdots)$, where $B_{\rm N}$ is a constant (Newtonian limit), gives $\d\ln B/\d\ln \x\approx k_1\x+(2k_2-k_1^2)\x^2+\cdots$.
A $s^2$ with $s_{\rm c}^2$ smaller than $4\x/3$ would then require $k_1<0$. However, this condition cannot be satisfied, since strong-field gravity in GR tends to reduce the stellar radius relative to its Newtonian value. Because $\widehat{R}\approx[\x/B(\x)]^{1/2}$ must decrease, one necessarily has $k_1>0$.
The central matter also cannot be conformal, as indicated by $s_{\rm c}^2/\x\to1$. In such a case, one would have
\begin{equation}\label{Rinf}
\frac{\d\ln\x}{\d\ln R}=0,~\text{ or }~
\frac{\d\ln R}{\d\ln\x}=\pm\infty.
\end{equation}
This implies that an infinitesimal change in $\x$ would induce an unbounded response in the radius $R$, i.e., the NS radius becomes infinitely sensitive to $\x$, which is clearly unphysical.

\section{A Gedankenexperiment and Physical Information Encapsulated in the Coefficient $A(\x)\equiv -a_2(\x)$}\label{SEC_a2}

The expression in Eq.\,(\ref{def-sc2BB}) is particularly useful, as the coefficient $B(\x)=-b_2(\x)$ effectively encodes the properties of $s_{\rm c}^2$. In this section, we examine the physical insights contained in the expansion coefficient $A(\x)\equiv -a_2(\x)$ by designing the following Gedankenexperiment.

\renewcommand*\figurename{\small FIG.}
\begin{figure}[h!]
\centering
\includegraphics[width=6.cm]{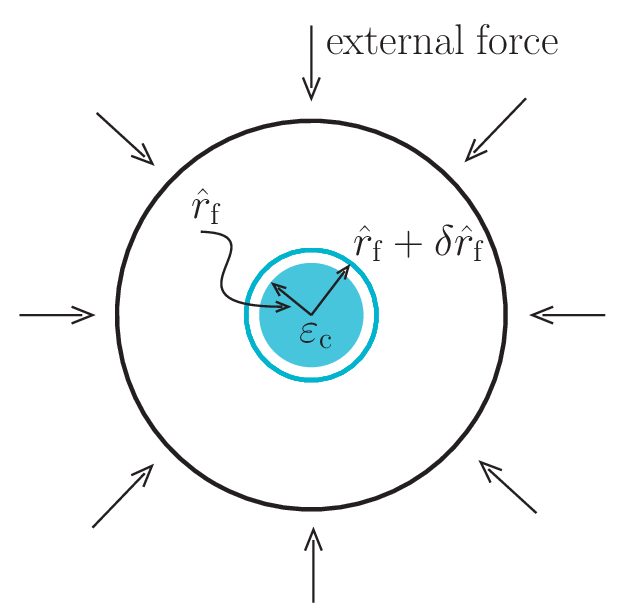}
\caption{(Color Online). A Gedankenexperiment: Exerting an external force (pressure) on an NS while keeping its central energy density $\varepsilon_{\rm c}$ fixed (therefore the $\heps(\hr)$ profile changes). The pressure cannot increase without bound; once the configuration turns unstable, the transition defines the upper limit of $\x = P_{\rm c}/\varepsilon_{\rm c}$.}\label{fig_a2Phys}
\end{figure}

\begin{figure}[h!]
\centering
\includegraphics[width=8.cm]{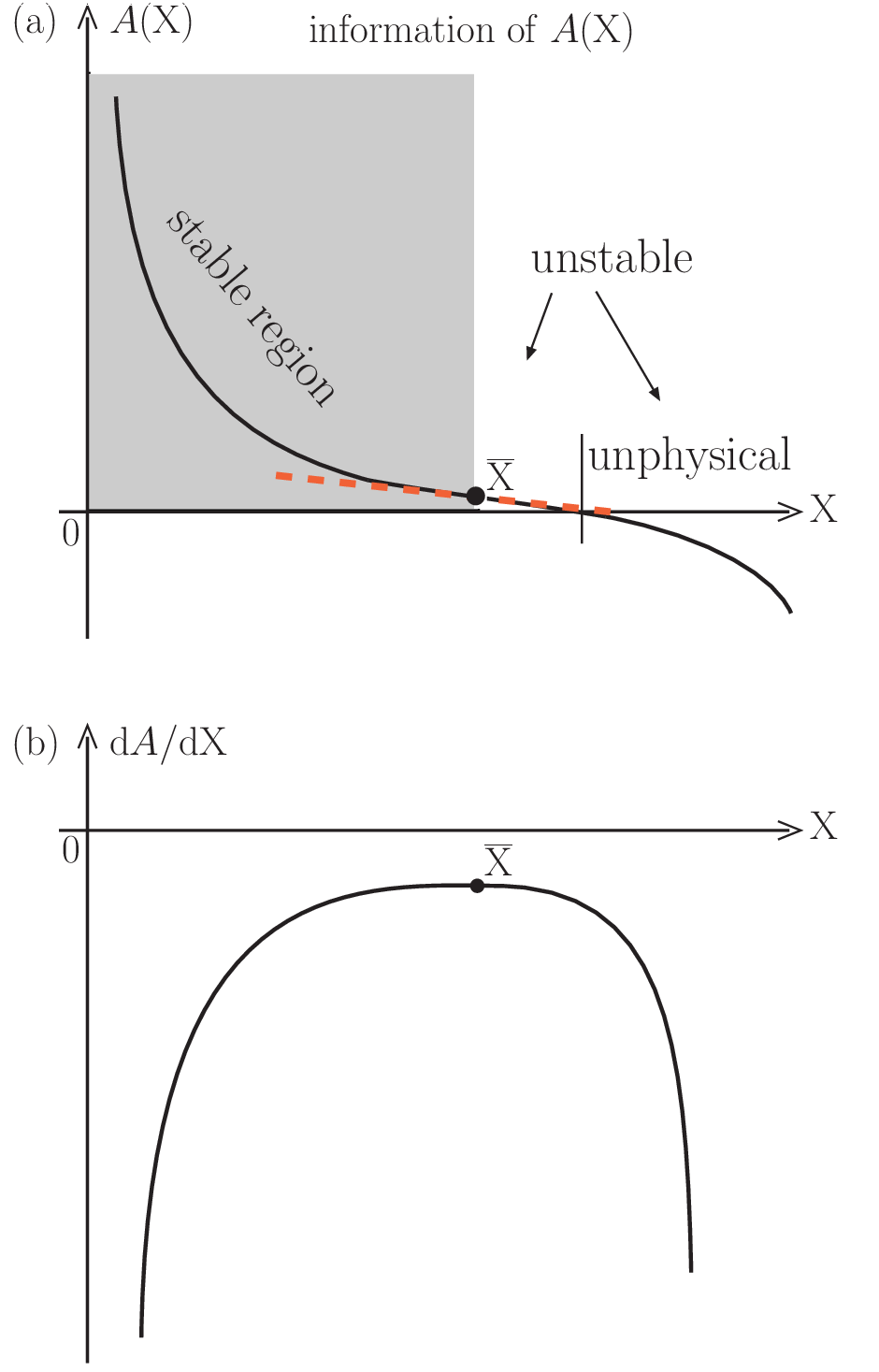}
\caption{(Color Online).
Information encoded in the coefficient $A(\x)$.
For small $\x$, the mass-sphere $\widehat{M}(\widehat{r}_{\rm f})$ with $\hr_{\rm f}$ fixed increases with $\x$, implying $\mathrm{d}A/\mathrm{d}\x < 0$ (panel (a)). 
As $\x$ grows, further increase of $\widehat{M}(\widehat{r}_{\rm f})$ (i.e., further compression at fixed $\widehat{r}_{\rm f}$ and $\varepsilon_{\rm c}$) becomes progressively more difficult. 
When $\x$ approaches a critical value $\overline{\x}$, the increase of $\widehat{M}(\widehat{r}_{\rm f})$ also reaches a corresponding critical value. 
For $\x > \overline{\x}$, the increase of $\widehat{M}(\widehat{r}_{\rm f})$ starts to accelerate again, meaning that compression becomes easier as $\x$ increases, indicating instability. 
The second-order derivative of $A(\x)$ with respect to $\x$ changes sign from positive to negative (panel (b)).
If $\x$ is even large then $A(\x)$ (or equivalently $a_2(\x)$) may become negative (positive) which is naturally unphysical.
}\label{fig_Ask}
\end{figure}

Suppose we have a NS with a fixed central energy density $\varepsilon_{\rm{c}}$ and then apply an external force to the NS, as shown in FIG.\,\ref{fig_a2Phys}.  
As this force increases, the central EOS-parameter $
\x= \widehat{P}_{\rm{c}}={P_{\rm{c}}}/{\varepsilon_{\rm{c}}}$ correspondingly increases.
Consider a sphere with a small fixed radius $\widehat{r}_{\rm f}$, or equivalently the mass-shell within $\delta\widehat{r}_{\rm f}$ near the center.  
For sufficiently small $\widehat{r}_{\rm f}$ ($\widehat{r}_{\rm f}\to0$), the expression for the small-sphere mass, $
\widehat{M}(\widehat{r}_{\rm f}) \approx 3^{-1}\widehat{r}_{\rm f}^3 + 5^{-1}a_2 \widehat{r}_{\rm f}^5$, becomes increasingly accurate, indicating that the coefficient $a_2$ determines the behavior of the small-sphere mass.  
Equivalently, $a_2$ controls the behavior of the corresponding mass-shell:
\begin{empheq}[box=\fbox]{align}
&\delta \widehat{M}(\widehat{r}_{\rm f})\approx\left(1+a_2\widehat{r}_{\rm f}^2\right)\widehat{r}_{\rm f}^2\delta\widehat{r}_{\rm f}\notag\\
\leftrightarrow&
a_2\ \text{determines}\ \delta \widehat{M}(\widehat{r}_{\rm f})\notag\\
&\hspace{1.cm} \text{for}\ \widehat{r}_{\rm f}\to0\ \text{and}\ \delta\widehat{r}_{\rm f}\approx\rm{fixed}.
\end{empheq}
Since in our analysis of this section, we consider the critical transition (singularity) near the very center $\hr_{\rm f}\approx0$, keeping the $a_2$-term is sufficient, as the next-order contribution is vanishingly small because $(a_4\hr_{\rm f}^4)/(a_2\hr_{\rm f}^2)=(a_4/a_2)\hr_{\rm f}^2$, considering that $a_4/a_2\sim\mathcal{O}(1)$ \cite{CL25-a}; namely the $a_2$-term is the only nontrivial contribution near $\hr_{\rm f}\approx0$, and the physical information it contains is meaningful.

Using $A(\x)$, we can rewrite the mass-shell as 
\begin{equation}\label{def-MS}
\delta \widehat{M}(\widehat{r}_{\rm{f}})\approx\left[1-A(\x)\widehat{r}_{\rm{f}}^2\right] \widehat{r}_{\rm{f}}^2\delta\widehat{r}_{\rm{f}},~~A(\x)>0.
\end{equation}
Since $\varepsilon_{\rm{c}}$ is fixed in our Gedankenexperiment, the physical mass of the small-sphere, $M(\widehat{r}_{\rm{f}})\sim\varepsilon_{\rm{c}}^{-1/2}\widehat{M}(\widehat{r}_{\rm{f}})$, or equivalently the physical mass of the shell, $\delta M(\widehat{r}_{\rm{f}})\sim\varepsilon_{\rm{c}}^{-1/2}\delta \widehat{M}(\widehat{r}_{\rm{f}})$, is essentially determined by $\delta \widehat{M}(\widehat{r}_{\rm{f}})$.  
We establish the general features of the coefficient $A(\x)$ as follows:
\begin{enumerate}[label=(\alph*),leftmargin=*]
\item As $\x$ increases, the physical mass $M(\widehat{r}_{\rm{f}})\sim\varepsilon_{\rm{c}}^{-1/2}\widehat{M}(\widehat{r}_{\rm{f}})$, or equivalently $\widehat{M}(\widehat{r}_{\rm{f}})$, also increases.  
This implies that $A(\x)$ is a decreasing function of $\x$, i.e., $\d A/\d\x<0$.

\item The increase of $\widehat{M}(\widehat{r}_{\rm{f}})$ (compression of the matter, since $\widehat{r}_{\rm{f}}$ and $\varepsilon_{\rm{c}}$ are fixed) becomes progressively more difficult as $\x$ increases, so the second-order derivative is positive for small $\x$: $\d^2A/\d\x^2>0$.
When $\x$ approaches some critical value $\overline{\x}$, the increase of $\widehat{M}(\widehat{r}_{\rm{f}})$ reaches a corresponding critical value.  
If $\x$ is pushed to even larger values, the rise of $\widehat{M}(\widehat{r}_{\rm f})$ accelerates rapidly. In physical terms, the system becomes increasingly ``willing'' to expand, revealing a mechanical instability of the configuration, what we refer to as the {\it mass-sphere} or {\it mass-shell instability.}
This behavior implies that $\x$ can not increase without limit (besides the naive limit 1), and correspondingly the second-order derivative of $A(\x)$ with respect to $\x$ transitions from positive to negative.  
The critical point $\overline{\x}$ is determined by the condition:
\begin{equation}
\boxed{
\text{onset of mass-sphere instability: }
\left.\frac{\d^2A}{\d\x^2}\right|_{\x=\overline{\x}}=0.}
\end{equation}
For $\x<\overline{\x}$, the mass evolution is stable, while for $\x>\overline{\x}$, it is unstable, characterized by an accelerated growth of the enclosed mass.
\end{enumerate}

These two criteria (a) and (b) can be summarized as follows: (i) mass should increase with $\x$ for small $\x$: ${\d \delta \widehat{M}(\widehat{r}_{\rm{f}})}/{\d\x}>0
\leftrightarrow{\d A}/{\d\x}<0$; (ii) mass becomes eventually harder to increase with $\x$: $
{\d ^2\delta \widehat{M}(\widehat{r}_{\rm{f}})}/{\d\x^2}<0
\leftrightarrow{\d ^2A}/{\d\x^2}>0$; and (iii) the mass-sphere instability occurs when $
{\d ^2\delta\widehat{M}(\widehat{r}_{\rm{f}})}/{\d\x^2}>0
\leftrightarrow{\d ^2A}/{\d\x^2}<0$ for $\x>\overline{\x}$, where $\widehat{r}_{\rm{f}}$ and $\varepsilon_{\rm{c}}$ are fixed and $\delta\widehat{M}(\widehat{r}_{\rm{f}})$ is given by Eq.\,(\ref{def-MS}).
See FIG.\,\ref{fig_Ask} for a sketch of the $A(\x)$ and $\d A/\d\x$ as functions of $\x$.
In the Newtonian limit, we have $s_{\rm{c}}^2(\x)\sim\x$, $A(\x)\sim \x^{-1}$, and therefore $
{\d^2A}/{\d\x^2}\sim\x^{-3}$, which is always positive; hence no such critical $\overline{\x}$ exists in this case.
It should be emphasized that the mass-sphere instability defined here is a concept within the IPAD-TOV framework. The condition $\d^2A/\d\x^2|_{\x=\overline{\x}}=0$ should therefore be regarded as an effective stability diagnostic, complementary to, rather than replacing, the conventional radial-mode or turning-point stability criteria.

The Principle of Causality, $s_{\rm{c}}^2 \leq 1$, also defines a critical value for $\x$, denoted as $\x_+$:
\begin{equation}\label{e2}
\boxed{
\text{causality boundary: }
s_{\rm{c}}^2\left(\x=\x_+\right)=1.}
\end{equation}
These two upper limits, $\x_+$ and $\overline{\x}$, are necessarily distinct; however, physically they should be close to each other, as both reflect the maximum compressibility allowed by the underlying physics of the NS cores.
We emphasize that even if $s_{\rm{c}}^2(\overline{\x})>1$, as often occurs in certain non-relativistic model calculations, the NS M-R relation still behaves reasonably with no strange features.  
This indicates that $\x_+$ and $\overline{\x}$ represent two independent criteria for setting the EOS-parameter upper bound.
More specifically, since the expression for $A(\x)$ for TOV NSs is
\begin{equation}
A(\x)=\frac{B(\x)}{s_{\rm{c}}^2(\x)}=\left.\frac{B(\x)}{\x}\left( 3-3\frac{\d\ln B}{\d\ln\x}\right)\right/\left( 4-3\frac{\d\ln B}{\d\ln\x}\right),
\end{equation}
see Eq.\,(\ref{def-sc2BB}),
the condition $\d^2A(\x)/\d\x^2=0$ reads
\begin{equation}
s_{\rm{c}}^4(\x)\frac{\d^2B}{\d\x^2}-2s_{\rm{c}}^2(\x)\frac{\d B}{\d\x}\frac{\d s_{\rm{c}}^2}{\d\x}
-s_{\rm{c}}^2(\x)B(\x)\frac{\d^2s_{\rm{c}}^4}{\d\x^2}+2B(\x)\left(
\frac{\d s_{\rm{c}}^2}{\d\x}\right)^2=0,
\end{equation}
which becomes (at the causality boundary $s_{\rm{c}}^2(\x)=1$) as
\begin{equation}\label{e1}
\frac{\d^2B}{\d\x^2}-2\frac{\d B}{\d\x}\frac{\d s_{\rm{c}}^2}{\d\x}
-B(\x)\frac{\d^2s_{\rm{c}}^4}{\d\x^2}+2B(\x)\left(
\frac{\d s_{\rm{c}}^2}{\d\x}\right)^2=0.
\end{equation}
This condition is fundamentally different from the causality boundary constraint of Eq.\,(\ref{e2}).
In fact, Eq.\,(\ref{e2}) for $s_{\rm{c}}^2(\x)=1$ and Eq.\,(\ref{e1}) for $\d^2A/\d\x^2=0$ (under $s_{\rm{c}}^2(\x)=1$) can be explicitly written as
\begin{equation}\label{e2-a}
\frac{3-4\x}{1-\x}\frac{B(\x)}{3\x}-\frac{\d B}{\d\x}=0,
\end{equation}
as well as,
\begin{align}
&\left[-9\x^4B^2(\x)\left(\frac{\d B}{\d\x}\right)+12\x^3B^3(\x)\right]\left(\frac{\d^3B}{\d\x^3}\right)\notag\\
&+18\x^4B^2(\x)\left(\frac{\d^2B}{\d\x^2}\right)^2
+\left[27\x^5\left(\frac{\d B}{\d\x}\right)^3
-126\x^4B(\x)\left(\frac{\d B}{\d\x}\right)^2\right.\notag\\
&\hspace{1cm}\left.+144\x^3B^2(\x)\left(\frac{\d B}{\d\x}\right)
-48\x^2B^3(\x)\right]\left(\frac{\d^2B}{\d\x^2}\right)\notag\\
&-36\x^4\left(\frac{\d B}{\d\x}\right)^4+198\x^3B(\x)\left(\frac{\d B}{\d\x}\right)^3-378\x^2B^2(\x)\left(\frac{\d B}{\d\x}\right)^2
\notag\\
&+312\x B^3(\x)\left(\frac{\d B}{\d\x}\right)-96B^4(\x)=0,\label{e1-a}
\end{align}
respectively.  
Clearly, Eqs.\,(\ref{e1-a}) and (\ref{e2-a}) are independent conditions for evaluating the upper limit of $\x$: Eq.\,(\ref{e2-a}) depends only on the first-order derivative of $B(\x)$, while Eq.\,(\ref{e1-a}) involves derivatives up to third-order, reflecting higher-order effects in the compressibility and stability of the dense matter.  

As mentioned above, physically and consistently, the $\x_+$ from $s_{\rm{c}}^2=1$ and the $\overline{\x}$ from $\d^2A/\d\x^2=0$ should be close to each other:
\begin{equation}\label{def-cons}
\text{physical requirement: } \x_+\approx\overline{\x}.
\end{equation}
Due to the perturbative scheme, $\x_+$ and $\overline{\x}$ may not necessarily be close to each other in our analysis.
Therefore, if they are not close, this consistency can be used to refine the upper limit for $\x$, which forms the main content of the following sections.
At even larger $\x>\x_{\rm{unphys}}$, $A(\x)$ may become negative due to the singular behavior of $s_{\rm{c}}^2$ ($\to \pm \infty$), corresponding to an unphysical state, since $A(\x)$ should remain positive. FIG.\,\ref{fig_Afun} sketches these regions:
\begin{equation}
    \x_{\rm{unphys}}\gtrsim\x_+\approx\overline{\x},
\end{equation}
and we call the region for $\x\gtrsim\x_+\approx\overline{\x}$ the unstable region while that for $\x\gtrsim\x_{\rm{unphys}}$ the unphysical region.

\section{Calculations of the Effective Correction to Upper Bound for $\x$ and Verification from the NS Compactness Scaling}\label{SEC_EFFC}

In this section, we analyze the effective correction to the upper bound for the central EOS-parameter $\x$ by employing the consistency condition of (\ref{def-cons}).
We first consider the $s_{\rm c}^2$ for TOV NSs without the $a_2$-term\,\cite{CLZ23-a}. In this case, the causality limit gives $\x\leq\x_+\approx 0.374$.  
Applying the mass-sphere stability condition $\d^2A/\d\x^2=0$ yields $\overline{\x}\approx0.377$.  
The proximity between $\x_+$ and $\overline{\x}$ is encouraging: it indicates that the causality bound and the mass-sphere stability point impose mutually consistent and complementary constraints on the central EOS-parameter. This observation may also help explain the effectiveness of the IPAD-TOV approach in capturing the effective scalings of the mass, radius, and compactness of NSs\,\cite{CL25-a}, even when the perturbative expansions are truncated at relatively low orders.

Next, we include the $a_2$-term in the NS mass by replacing the central energy density $\widehat{\varepsilon}_{\rm{c}}=1$ with the average energy density $\langle\widehat{\varepsilon}\rangle\approx 1+a_2\widehat{r}^2$, which modifies the $s_{\rm c}^2$ to\,\cite{CL25-b}
\begin{equation}\label{ew-0}
s_{\rm{c}}^2(\x)=\x\left(1+\frac{1}{3}\frac{1+3\x^2+4\x}{1-3\x^2}\right)\left(1-\frac{3}{25}\x\right),
\end{equation}
while the expression for $B(\x)$ remains unchanged. In this case, the numerical values become $\x_+\approx0.381$ and $\overline{\x}\approx0.368$.  
Again, $\x_+$ and $\overline{\x}$ are close to each other, indicating the internal consistency of our approach and reinforcing the physical significance of these upper-bound criteria.  

For reference, the singularity of the $s_{\rm{c}}^2(\x)$ occurs at $\x_{\rm{unphys}}=1/\sqrt{3}\approx0.577$, representing an unphysical state (as indicated in FIG.\,\ref{fig_Ask}) where the $s_{\rm c}^2$ diverges, beyond which the model is no longer physically meaningful.

Going beyond the leading-order expansion of $\widehat{P}$ in the dimensionless radial coordinate
$\widehat{r}$ becomes eventually complicated, as it requires analytically solving the TOV equations together with a specified dense matter EOS. Nevertheless, the NS pressure profile can be written in the effective
form
\begin{equation}\label{def-Pfr}
\widehat{P}(\widehat{r})=\x - B\widehat{r}^{2} + f(\widehat{r}),
\end{equation}
where $f(\widehat{r})$ collects all higher-order contributions in $\widehat{r}$ starting at
$\mathcal{O}(\widehat{r}^{4})$. Evaluating Eq.\,(\ref{def-Pfr}) at $\widehat{r}=\widehat{R}$ then
determines the NS radius through the condition
$\x - B\widehat{R}^{2} + f(\widehat{R}) = 0$.
The explicit form of $f(\widehat{r})$ is uniquely fixed by the TOV equations together with the
dense matter EOS; in this sense, Eq.\,(\ref{def-Pfr}) remains exact.
However, although the exact expansion of $\widehat{P}(\widehat{r})$ contains only higher-order
corrections $\sim \widehat{r}^{4}, \widehat{r}^{6}, \cdots$, the NS radius is determined by the pressure at a finite $\widehat{r}$, where the cumulative impact of these terms becomes important. To effectively model
the complexity of $f(\widehat{r})$ without committing to a specific microscopic EOS, we therefore
introduce a dimensionless parameter $\sigma$ and adopt the effective parametrization
\begin{equation}
f(\widehat{r}) = -\sigma\x B\widehat{r}^{2},
\end{equation}
where $\sigma$ is to be determined self-consistently.
From this viewpoint, the quadratic correction should be interpreted as an effective renormalization
of the coefficient of the leading-$\widehat{r}^{2}$ term, arising from the cumulative influence of microphysical interactions and nonlinear gravitational effects. In this sense, the parameter $\sigma$ encodes how subleading contributions are projected onto the dominant curvature
term of $\widehat{P}(\widehat{r})$ on macroscopic scales. A quadratic form therefore represents the
natural leading effective choice, since the $B$-term already multiplies $\widehat{r}^{2}$ and
symmetry requires that only even powers of $\widehat{r}$ appear in the expansion of
$\widehat{P}$\,\cite{CL24-a}; correspondingly, the genuine subleading corrections originate from
terms such as $\widehat{r}^{4}, \widehat{r}^{6}, \cdots$.

In the next section, we show that adopting alternative functional forms for $f(\widehat{r})$ within this framework leads to very similar constraints on the upper bound of $\x$, demonstrating the robustness of our conclusions.
This is reasonable and understandable: even without the effective $f$-correction, the two bounds on $\x$ are already close, indicating that the $f$-correction is a perturbation.
In addition, for small $\x$, the correction $\sigma\x \to 0$, i.e., it has no impact on low-mass NSs. 
The effective $B$ parameter then becomes
\begin{equation}
B(\x)\to B_{\rm{eff}}(\x)=B(\x)\left(1+\sigma\x\right)=\frac{1}{6}\left(1+3\x^2+4\x\right)\left(1+\sigma \x\right).
\end{equation}
Consequently, the dimensionless radius reads
\begin{equation}
\widehat{R}=\sqrt{\frac{6\x}{1+3\x^2+4\x}}\cdot\left(\frac{1}{1+\sigma\x}\right)^{1/2},
\end{equation}
and the physical radius becomes
\begin{equation}
R=
\frac{C_R}{\sqrt{\varepsilon_{\rm{c}}}}
\left(\frac{\x}{1+3\x^2+4\x}\right)^{1/2}\cdot\left(\frac{1}{1+\sigma\x}\right)^{1/2},
\end{equation}
where $C_R$ is a scaling constant.

The energy density expansion is modified to $\widehat{\varepsilon}(\widehat{r})\approx1+a_2^{\sigma}\widehat{r}^2+\cdots$, where the superscript ``$\sigma$'' marks the effective correction on the coefficient $a_2$, with $
a_2^{\sigma}=b_2(1+\sigma\x)/s_{\rm{c}}^2$, and thus
\begin{equation}
a_2^{\sigma}\widehat{R}^2=-{B_{\rm{eff}}\widehat{R}^2}/{s_{\rm{c}}^2}=-{\x}/{s_{\rm{c}}^2},
\end{equation}
showing that the final expression (``$-\x/s_{\rm c}^2$'') is identical to the $\sigma=0$ case.
The NS mass can be expressed as
\begin{align}\label{ew-1}
M_{\rm{NS}}\approx&\frac{1}{3}\widehat{R}^3\left(1+\frac{3}{5}a_2^{\sigma}\widehat{R}^2\right)Q
\approx\frac{1}{3}\widehat{R}^3\left(1-\frac{3}{5}\frac{\x}{s_{\rm{c}}^2}\right)Q\notag\\
\approx&
\frac{C_M}{\sqrt{\varepsilon_{\rm{c}}}}\left(\frac{\x}{1+3\x^2+4\x}\right)^{3/2}\cdot\left(\frac{1}{1+\sigma\x}\right)^{3/2}\cdot\left(1-\frac{3}{5}\frac{\x}{s_{\rm{c}}^2}\right),
\end{align}
where $C_M$ is another scaling constant.
Furthermore, the effective correction to $s_{\rm{c}}^2$ is introduced via a parameter ``$\kappa$'' as:
\begin{equation}\label{ew-3}
s_{\rm{c}}^2(\x)\approx\x\left(1+\frac{1}{3}\frac{1+3\x^2+4\x}{1-3\x^2}\right)\left(1-\frac{3}{25}\x\right)\left(1+\kappa\x\right),
\end{equation}
see Eq.\,(\ref{ew-0}).
Expanding for small $\x$ of Eq.\,(\ref{ew-3}) then gives
\begin{equation}\label{ew-4}
s_{\rm{c}}^2(\x)\approx\frac{4}{3}\x + \left(\frac{88}{75} + \frac{4\kappa}{3}\right)\x^2,
\end{equation}
and on the other hand requiring $\d M_{\rm{NS}}/\d\varepsilon_{\rm{c}}=0$ yields
\begin{equation}\label{ew-5}
s_{\rm{c}}^2(\x)\approx\frac{4}{3}\x + \left(\frac{88}{75} -\frac{2\kappa}{11}+\frac{\sigma}{3}\right)\x^2.
\end{equation}
Matching these two expansions for $s_{\rm c}^2$ leads to 
\begin{equation}
\kappa={11\sigma}/{50},
\end{equation}
so that the final effective $s^2$ reads
\begin{equation}\label{def-sc2eff}
s_{\rm{c}}^2(\x)\approx\x\left(1+\frac{1}{3}\frac{1+3\x^2+4\x}{1-3\x^2}\right)\cdot
\overbrace{
\left(1+\frac{11\sigma-6}{50}\x\right)}^{\mbox{with correction}},
\end{equation}
here only the linear term in $\x$ in the correction is kept.
This expression captures the combined influence of higher-order contributions from the expansion in the IPAD-TOV approach (the $a_2$-term and Eq.\,(\ref{ew-0})),  and the effective corrections introduced through the function $f(\hr)$, compared with Eq.\,(\ref{sc2-gen}) with $\Psi=0$.

\begin{figure}[h!]
\centering
\includegraphics[width=8.cm]{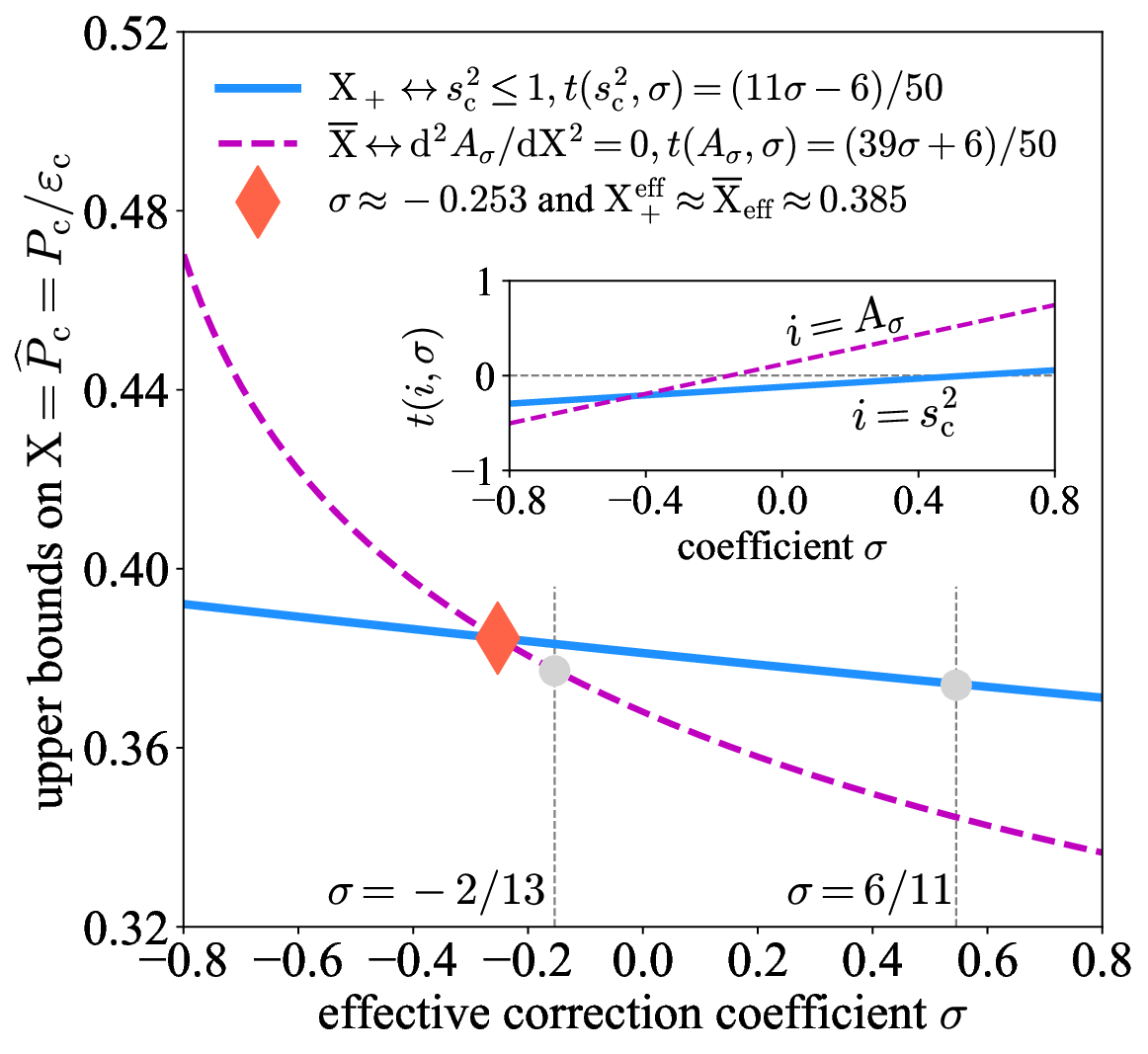}
\caption{(Color Online).
Dependence of $\x_+$ and $\overline{\x}$ on the coefficient $\sigma$. 
The intersection of the two curves gives the final estimate of the upper bound on $\x$. 
The inset shows the effective corrections to $s_{\rm c}^2$ and $A_{\sigma}(\x)$, and the two vertical dashed lines mark the positions at which the effective corrections in $A_{\sigma}(\x)$ and $s_{\rm c}^2$ vanish, respectively. 
See the text for details.
}\label{fig_seaching_sigma}
\end{figure}

\begin{figure}[h!]
\centering
\includegraphics[width=8.cm]{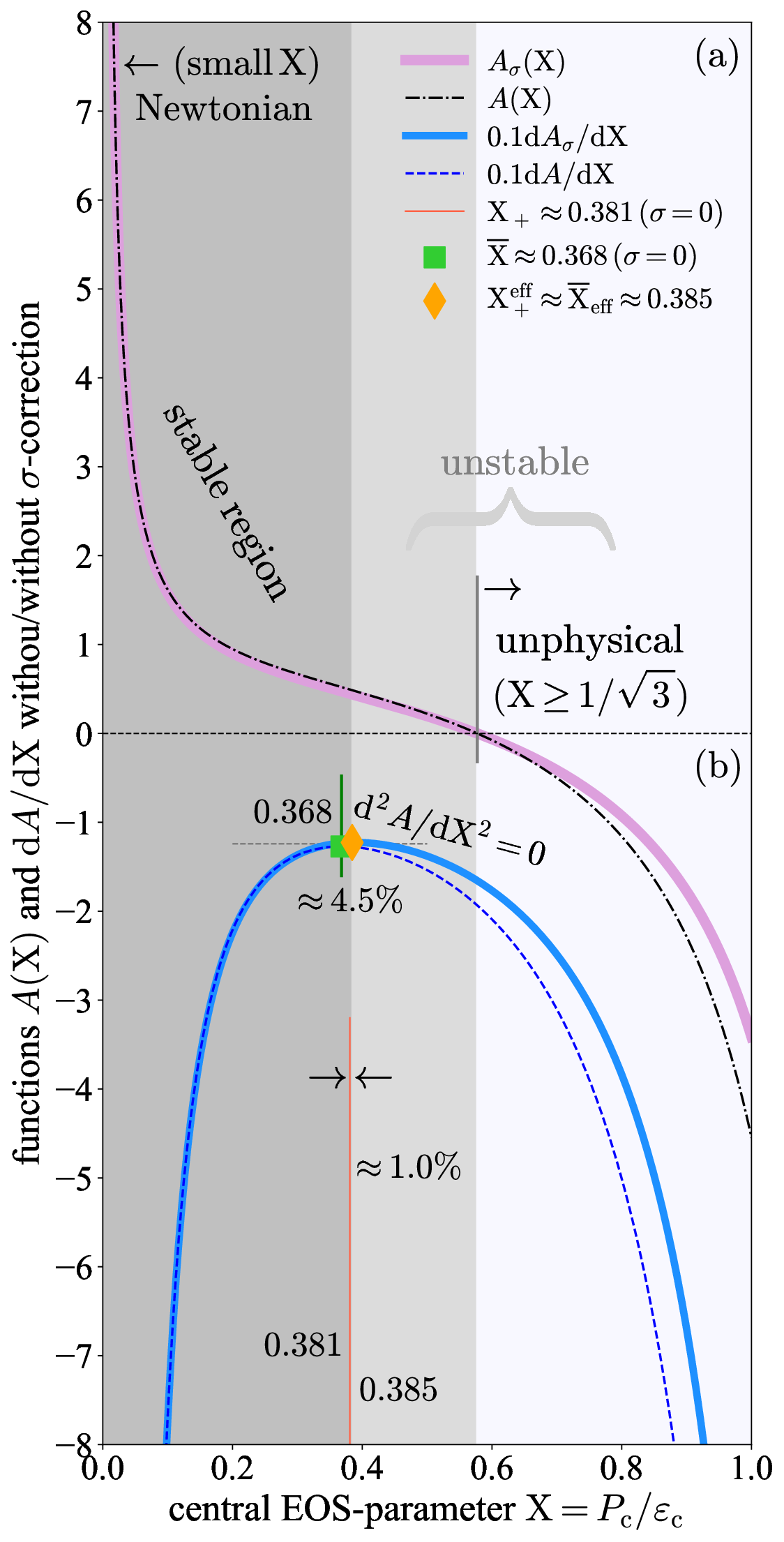}
\caption{(Color Online).
The numerical calculation corresponds to FIG.\,\ref{fig_Ask}. 
Without the $\sigma$-correction, $\overline{\x}$ and $\x_+$ are found to be about 0.368 and 0.381, respectively. 
Including the $\sigma$-correction shifts both values to $\approx 0.385$, inducing effects of about 4.5\% and 1.0\%, respectively. 
For $\x \ge 1/\sqrt{3} \approx 0.577$, the coefficient $A(\x)$ becomes negative, corresponding to an unphysical state.
}\label{fig_Afun}
\end{figure}

\begin{figure*}
\centering
\includegraphics[width=8.5cm]{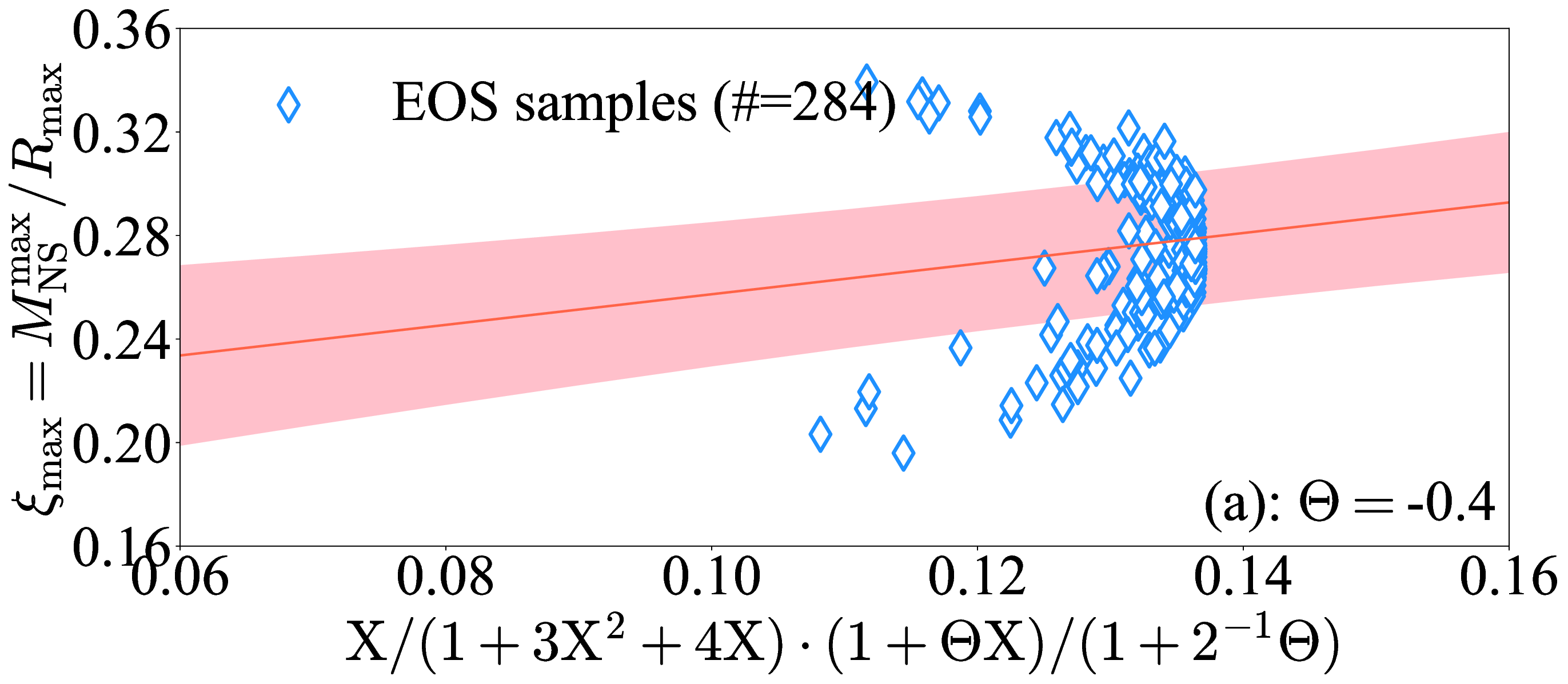}\quad
\includegraphics[width=8.5cm]{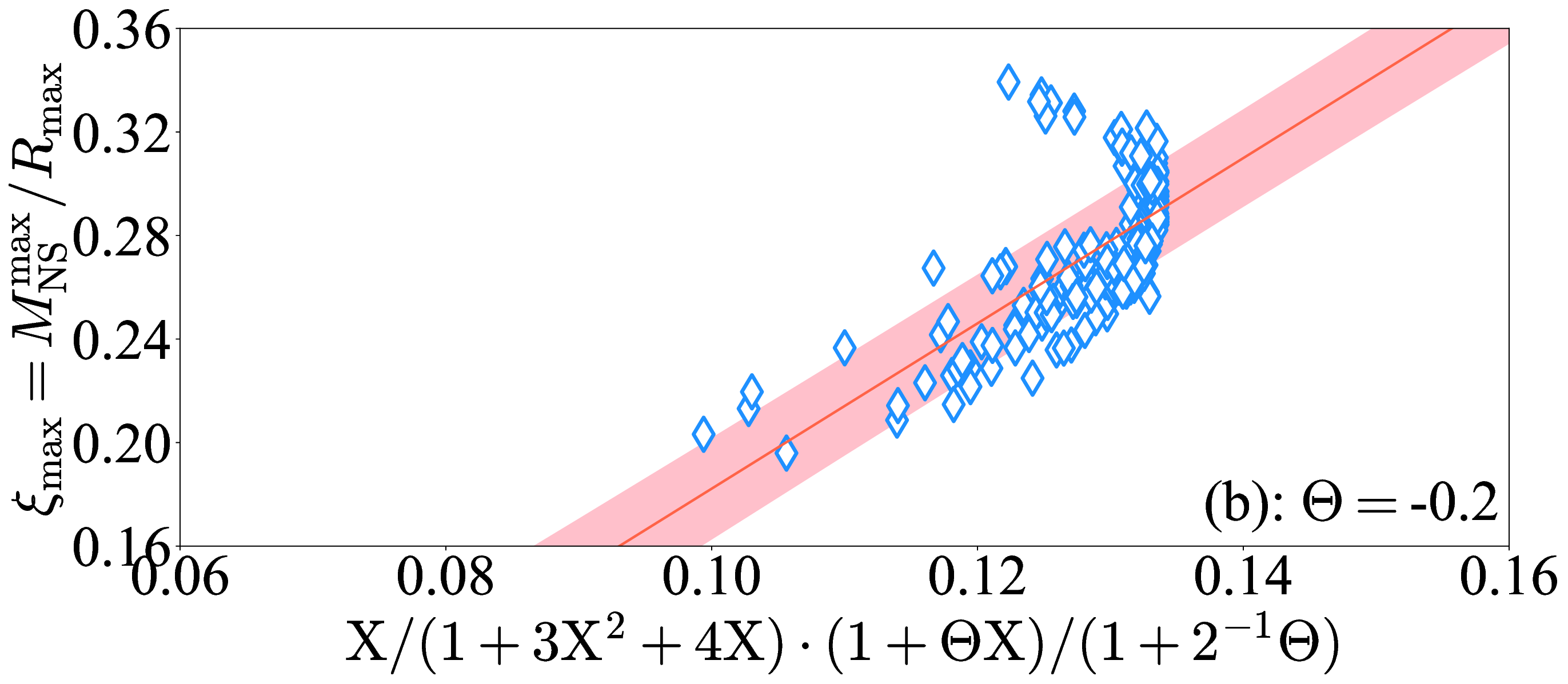}\\
\includegraphics[width=8.5cm]{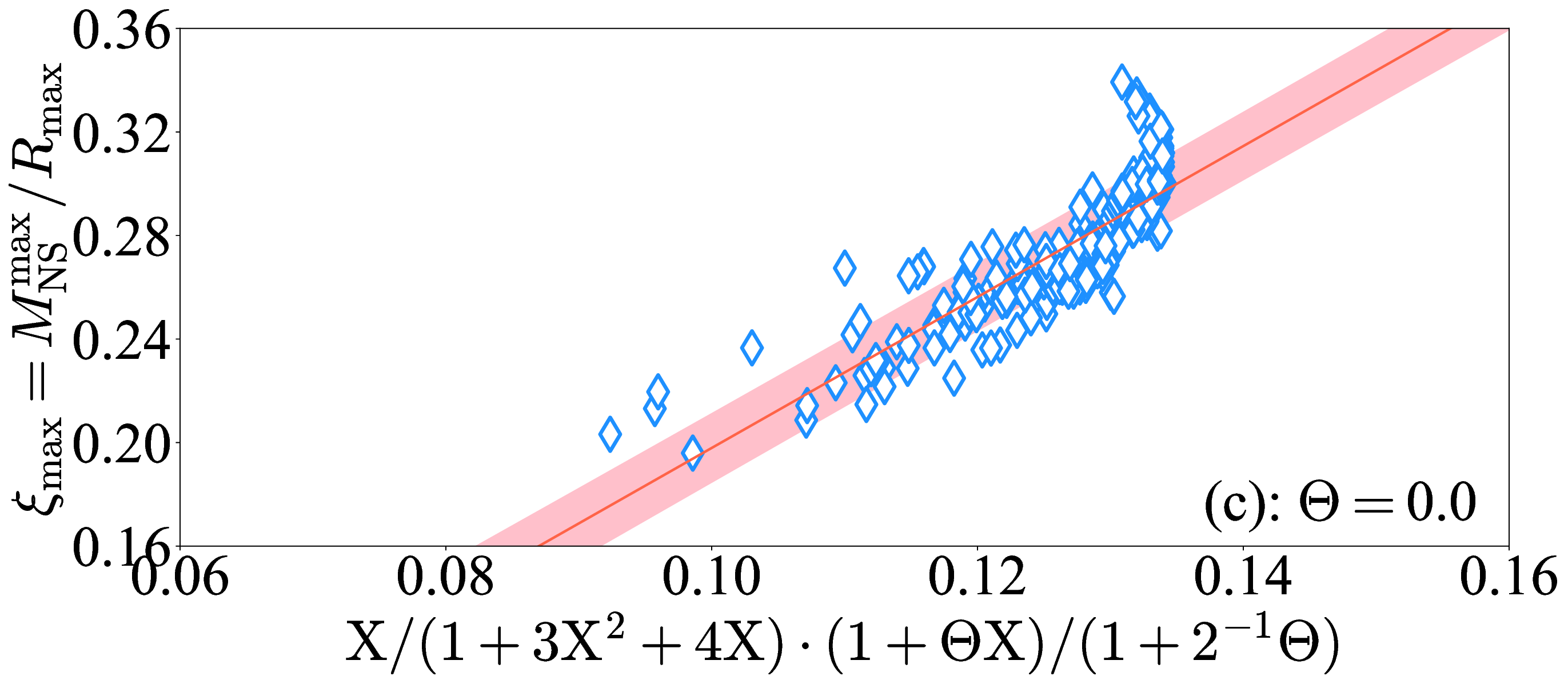}\quad
\includegraphics[width=8.5cm]{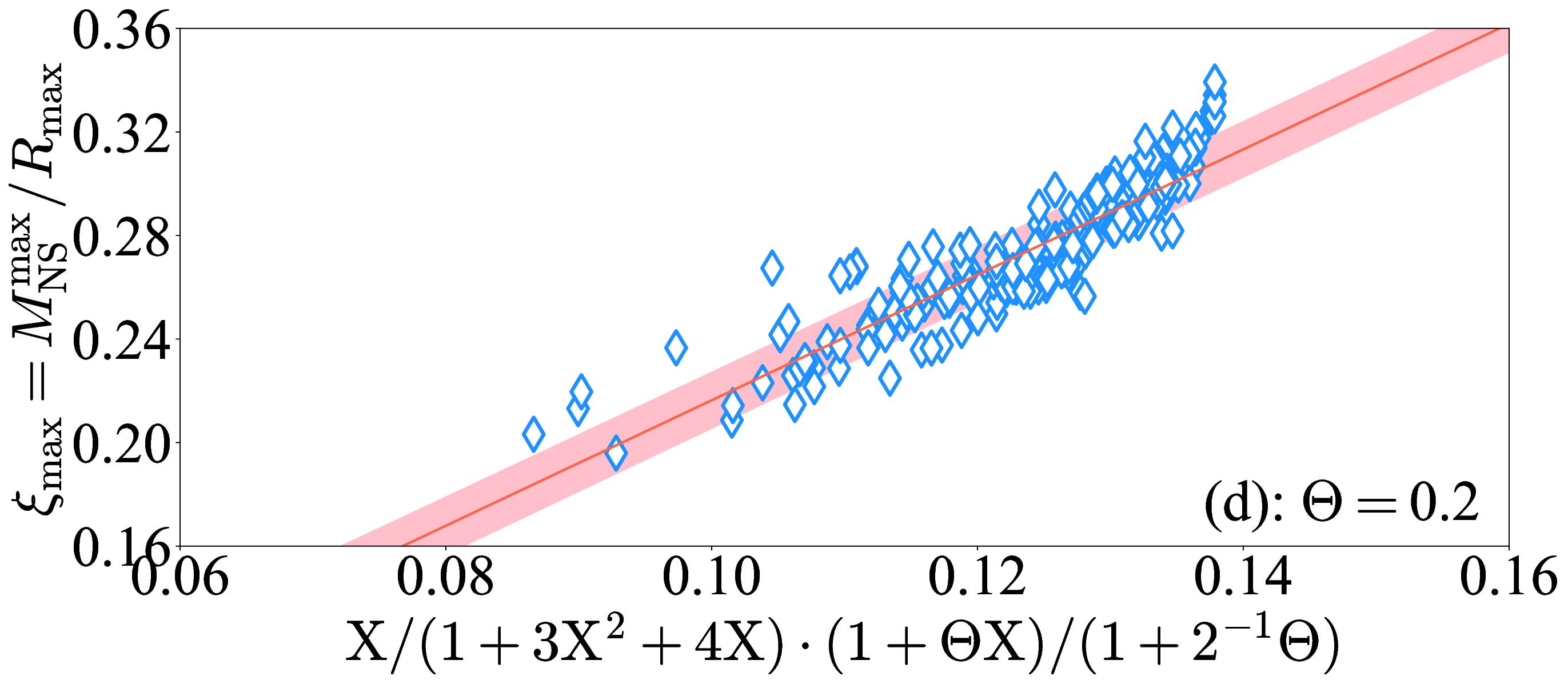}\\
\includegraphics[width=8.5cm]{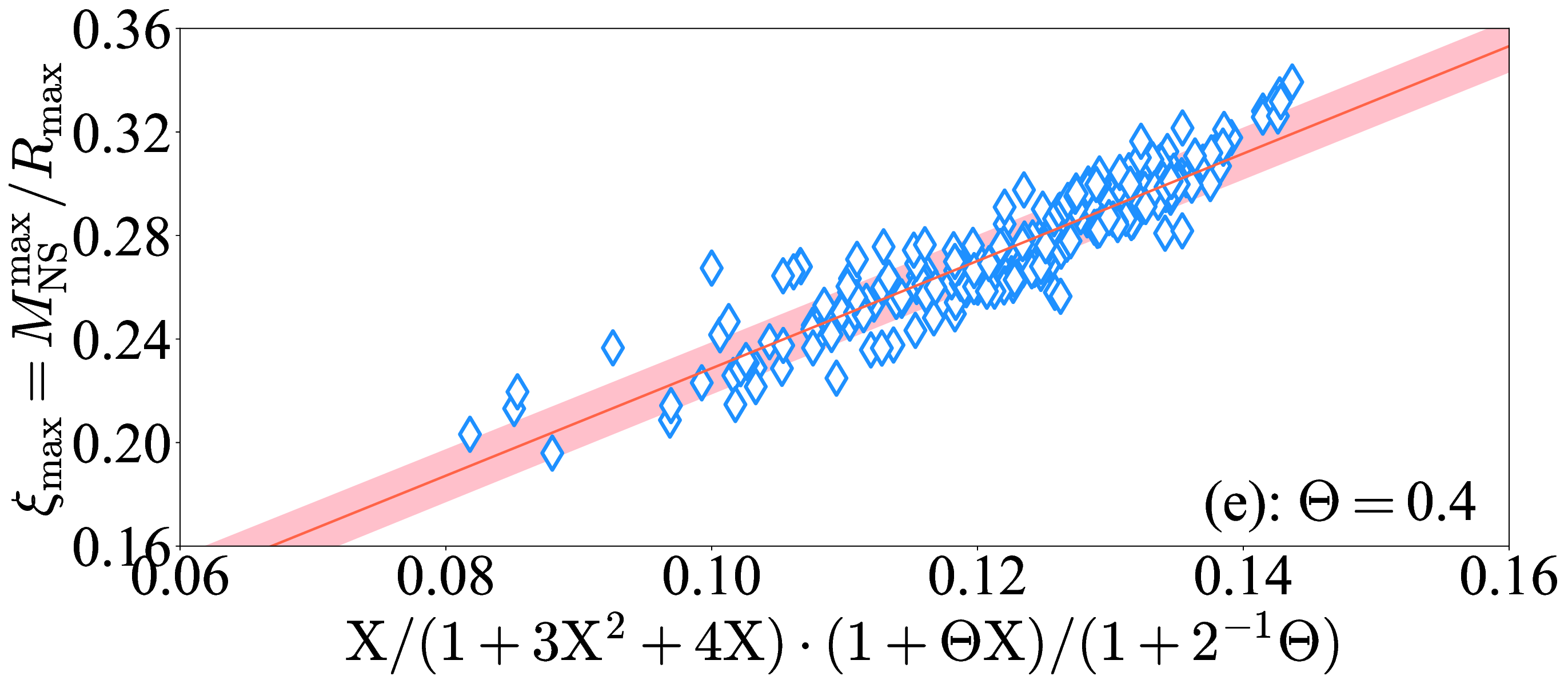}\quad
\includegraphics[width=8.5cm]{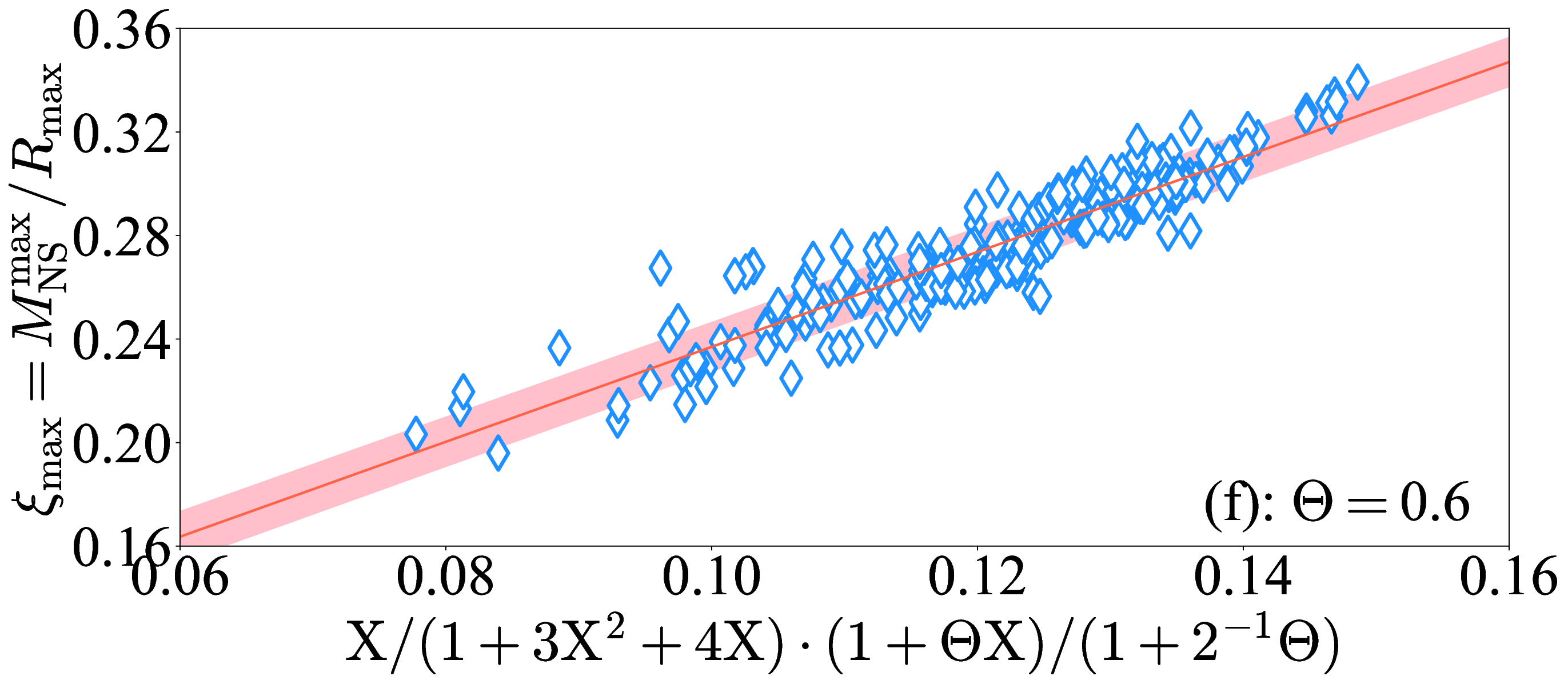}\\
\includegraphics[width=8.5cm]{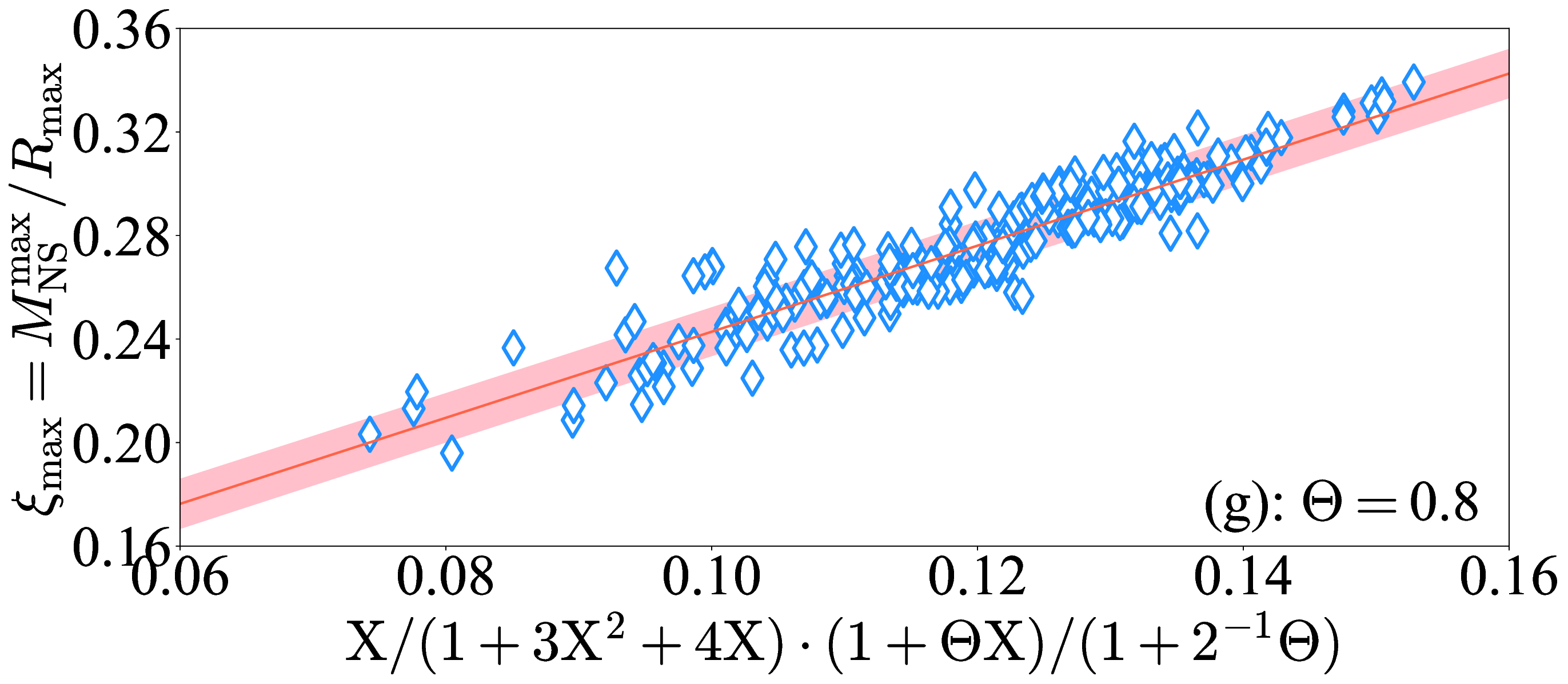}\quad
\includegraphics[width=8.5cm]{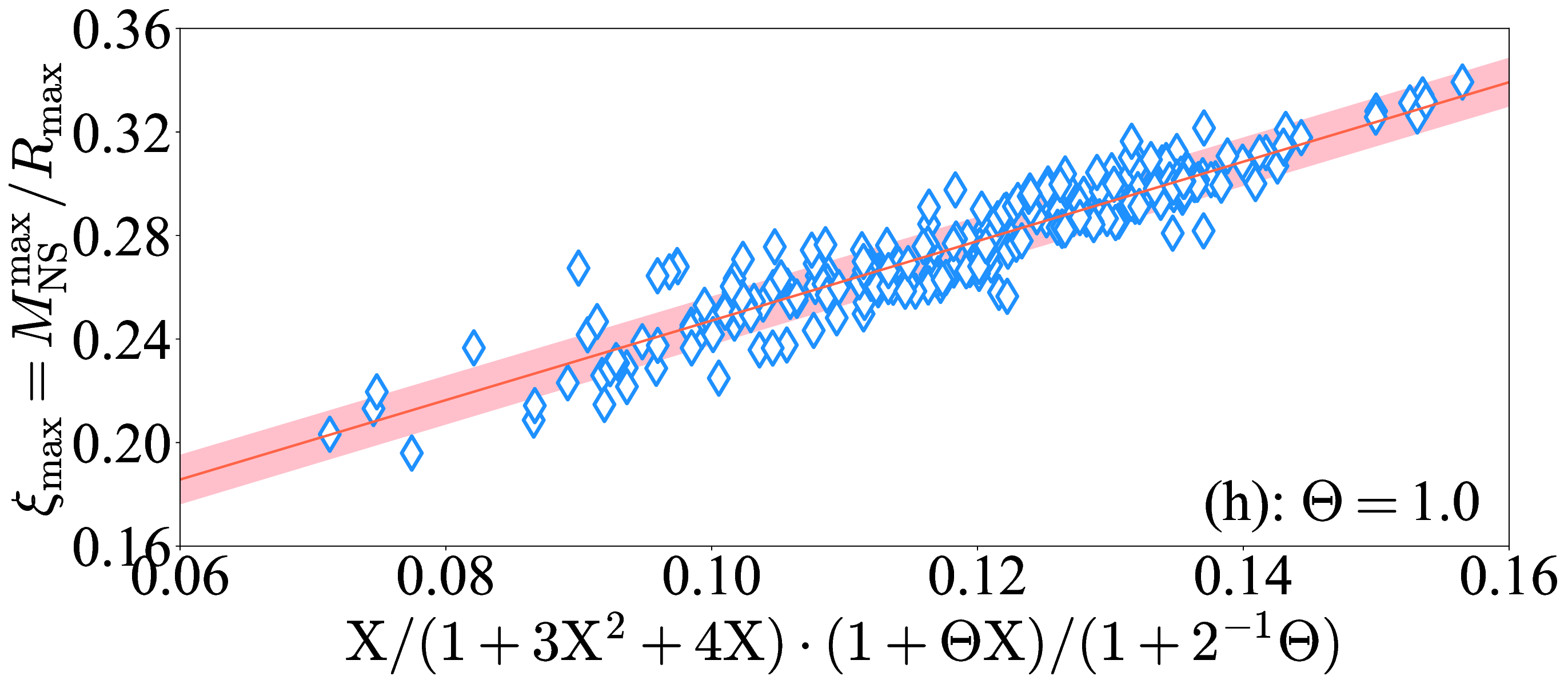}\\
\includegraphics[width=8.5cm]{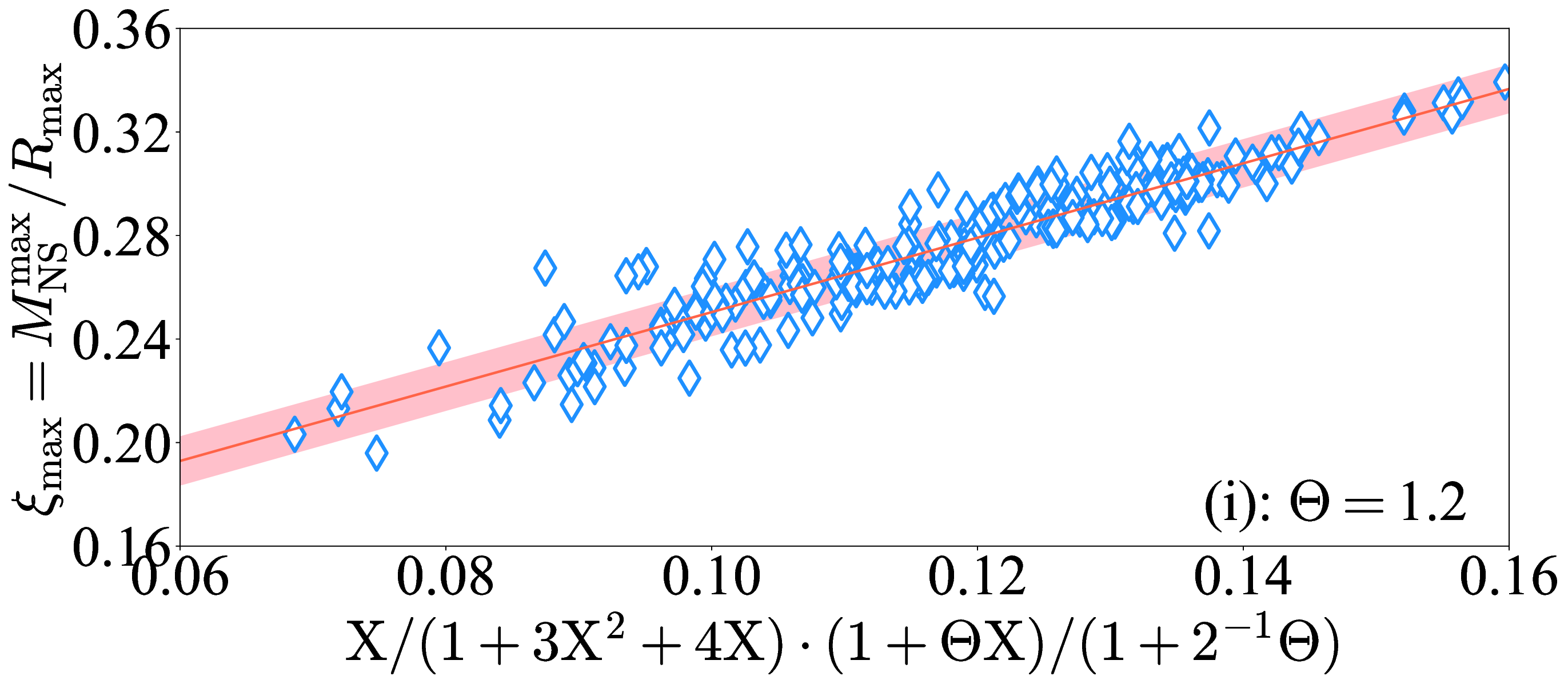}\quad
\includegraphics[width=8.5cm]{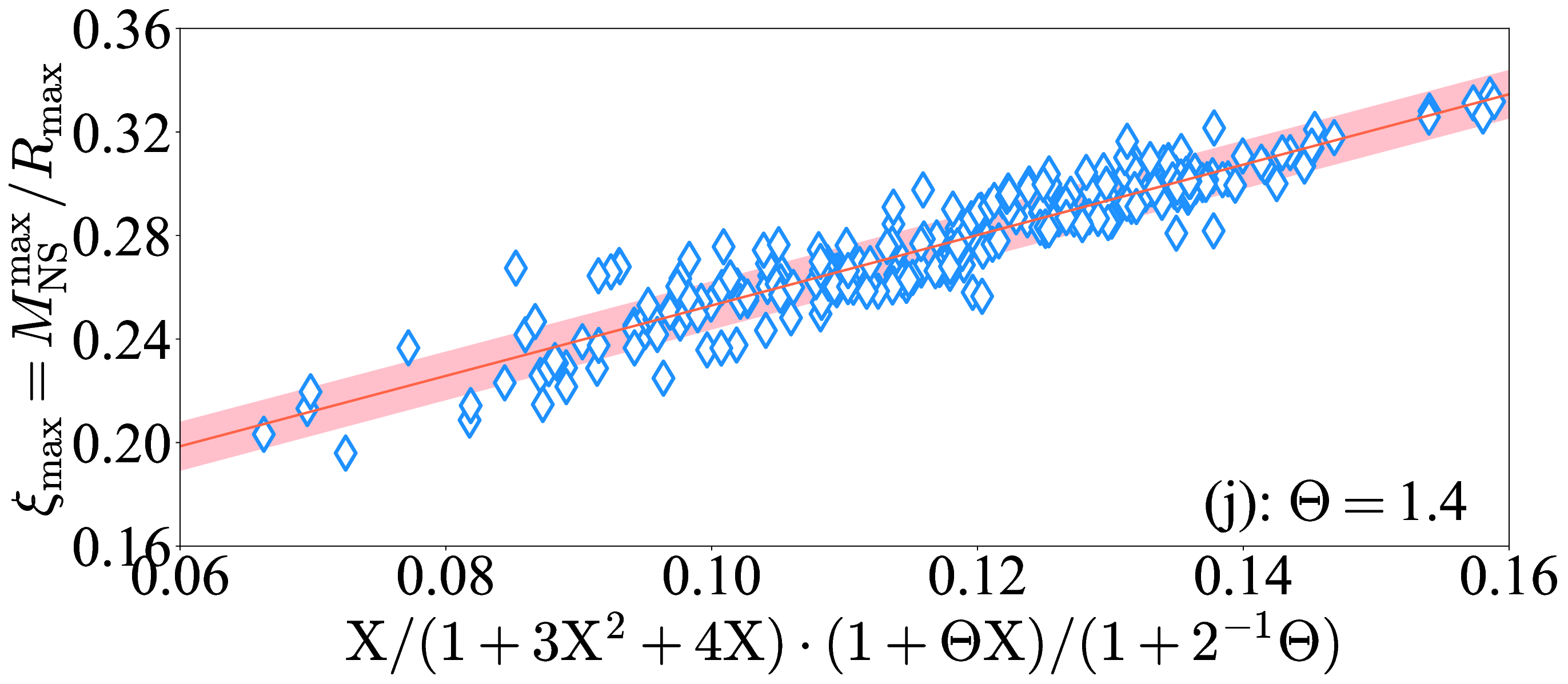}
\includegraphics[width=8.5cm]{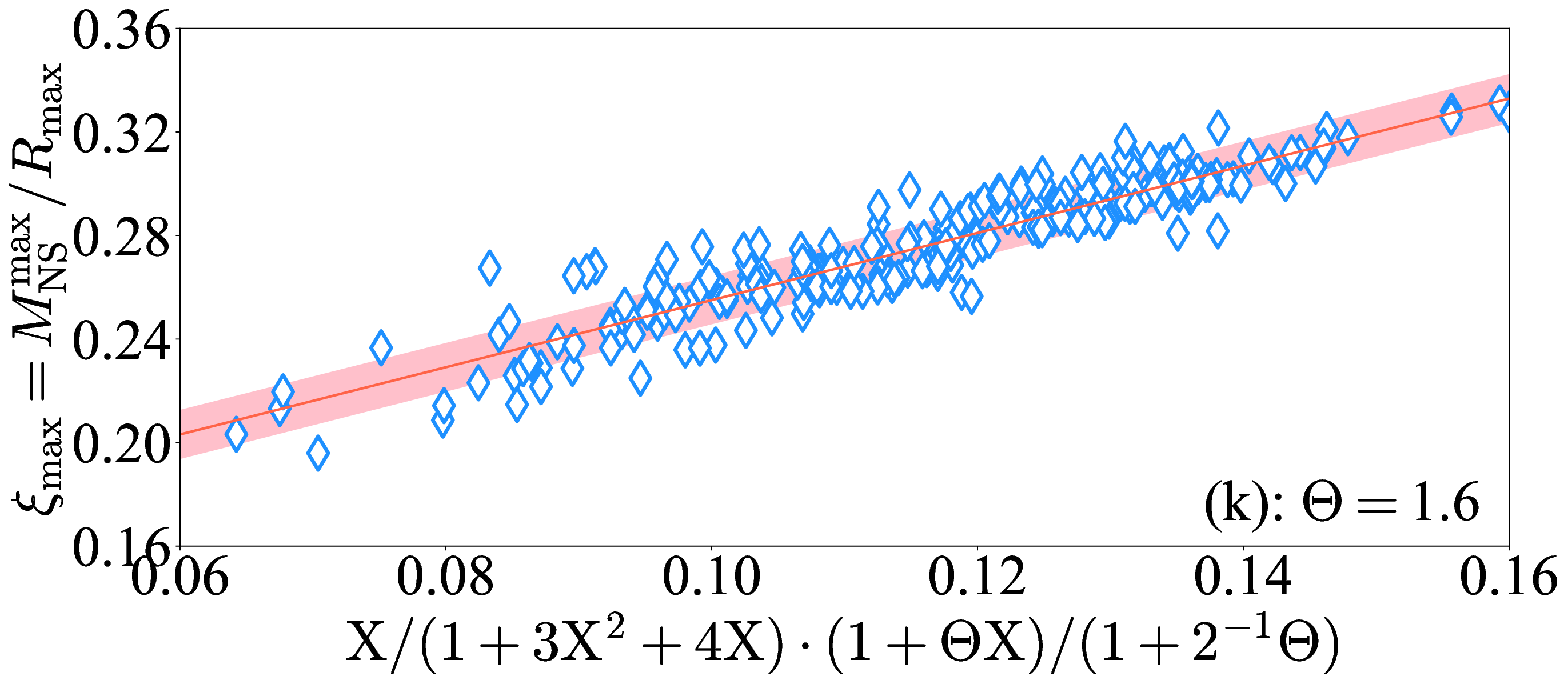}\quad
\includegraphics[width=8.5cm]{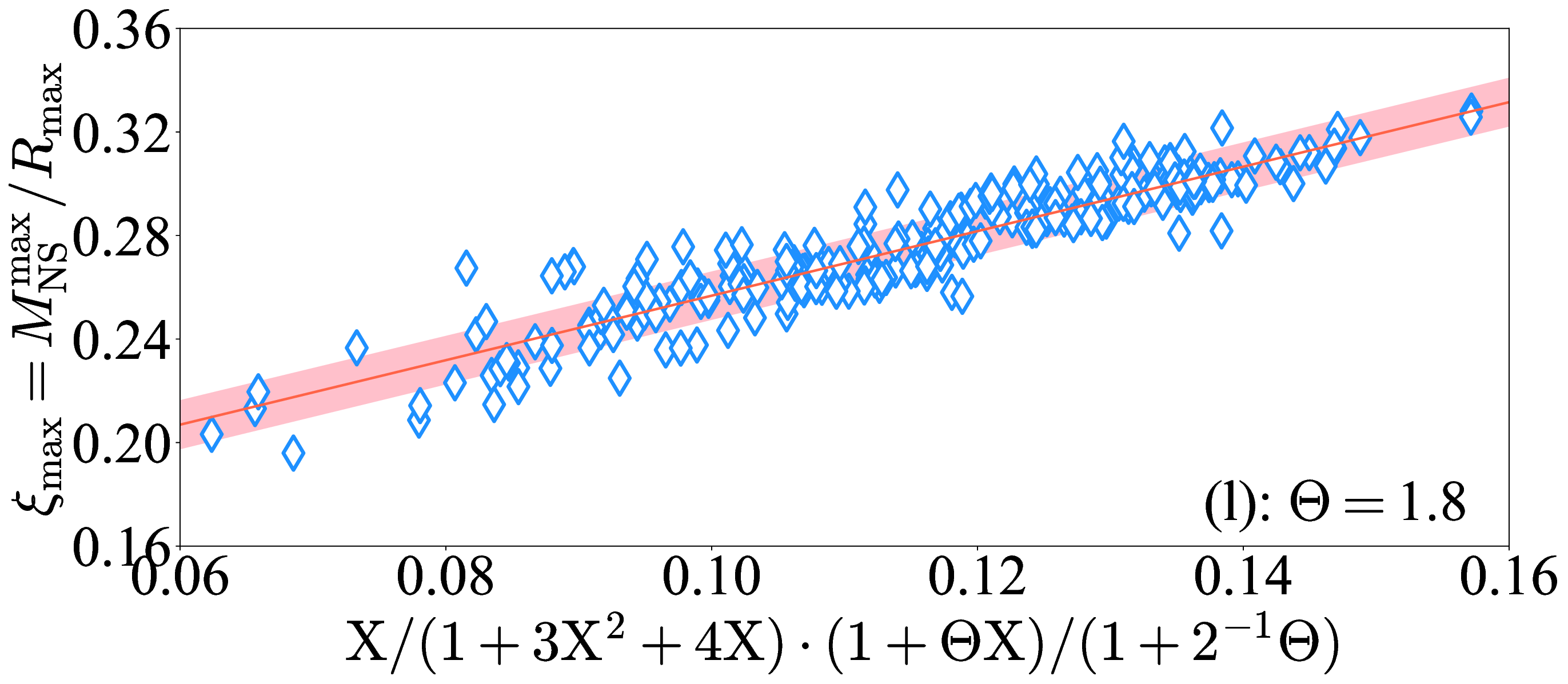}
\caption{(Color Online).
Impact of including the $\Theta$-term (fitting parameter) on the compactness scaling for NSs at the maximum-mass configuration. A total of 284 EOS samples are used, including those using microscopic calculations as well as relativistic and non-relativistic phenomenological models, with or without phase transitions or continuous crossover. 
See the text for details.
}\label{fig_rp_xi}
\end{figure*}

We now determine the value of the effective parameter $\sigma$.  
The function $A(\x)$ becomes (a subscript ``$\sigma$'' is added to indicate its dependence on $\sigma$):
\begin{align}\label{def-Aeff}
&A(\x)\to A_{\sigma}(\x)=\frac{B_{\rm{eff}}(\x)}{s_{\rm{c}}^2(\x)}
\approx\frac{1+3\x^2+4\x}{6\x}\notag\\
&\times
\left(1+\frac{1}{3}\frac{1+3\x^2+4\x}{1-3\x^2}\right)^{-1}\cdot\overbrace{\left(1+\frac{39\sigma+6}{50}\x\right)^{-1}}^{\mbox{with correction}},
\end{align}
where $B_{\rm{eff}}=B(1+\sigma\x)$ is used here and only the linear term in $\x$ in the correction is kept when calculating $B_{\rm{eff}}(\x)/s_{\rm{c}}^2$; this expression for $A_\sigma(\x)$ incorporates the effects of the correction $f(\widehat{r})$ through $\sigma$, which affects both $B_{\rm{eff}}(\x)$ and $s_{\rm{c}}^2(\x)$.  
The effective upper bounds $\x_+^{\rm{eff}}$ and $\overline{\x}_{\rm{eff}}$ can be determined from the conditions $s_{\rm{c}}^2(\x)=1$ and $\d^2A_{\sigma}/\d\x^2=0$, respectively.

By requiring consistency, i.e., $\x_+^{\rm{eff}}=\overline{\x}_{\rm{eff}}$, we obtain the correction parameter $\sigma\approx-0.253$ and simultaneously $\x_+^{\rm{eff}}\approx\overline{\x}_{\rm{eff}}\approx0.385$, as shown in FIG.\,\ref{fig_seaching_sigma}.  
In this figure, the two vertical dashed lines mark the positions where the effective corrections in $A_{\sigma}$ and $s_{\rm{c}}^2$ vanish (see Eqs.\,(\ref{def-Aeff}) and (\ref{def-sc2eff})), namely $\sigma=-2/13$ for $A_\sigma$ and $\sigma=6/11$ for $s_{\rm c}^2$.  
The corresponding upper bounds at these positions are 0.377 from $\d^2A_{\sigma}/\d\x^2=0$ and 0.374 from $s_{\rm{c}}^2=1$, shown by the solid grey circles, which were obtained earlier in this section.  
The inset illustrates the correction coefficients $t(i,\sigma)$ with $i$ taking $s_{\rm{c}}^2$ or $A_{\sigma}$, i.e., $t(s_{\rm{c}}^2,\sigma)=(11\sigma-6)/50$ and $t(A_{\sigma},\sigma)=(39\sigma+6)/50$, both of which are negative for $\sigma\approx-0.253$.  
The impact of the correction on the upper bound of the central EOS-parameter $\x$ due to the causality condition $s_{\rm{c}}^2\leq1$ is $\approx1.0\%$ (compared with 0.381), while the impact relative to the $\overline{\x}$ from mass-sphere stability condition is $\approx 4.5\%$ (compared with 0.368). 
Accordingly, the NS radius and mass scalings are modified as
\begin{align}
R\sim&\frac{\Pi_{\rm c}^{1/2}}{\sqrt{\varepsilon_{\rm{c}}}}\cdot\left(1-\frac{1}{2}\sigma\x\right),~~
M_{\rm{NS}}\sim\frac{\Pi_{\rm c}^{3/2}}{\sqrt{\varepsilon_{\rm{c}}}}\cdot\left(1+\frac{18-33\sigma}{25}\x\right),
\end{align}
and the NS compactness becomes
\begin{align}
\xi=\frac{M_{\rm{NS}}}{R}\sim\Pi_{\rm c}\cdot\left(1+\frac{36-41\sigma}{50}\x\right),
\end{align}
where $\sigma\approx-0.253$. 
The term in the bracket represents the effective correction to the compactness scaling, giving a factor of approximately $1+0.927\x$.  
FIG.\,\ref{fig_Afun} shows the corresponding quantities, which align with the sketch presented in FIG.\,\ref{fig_Ask}.

\begin{figure}[h!]
\centering
\includegraphics[width=8.cm]{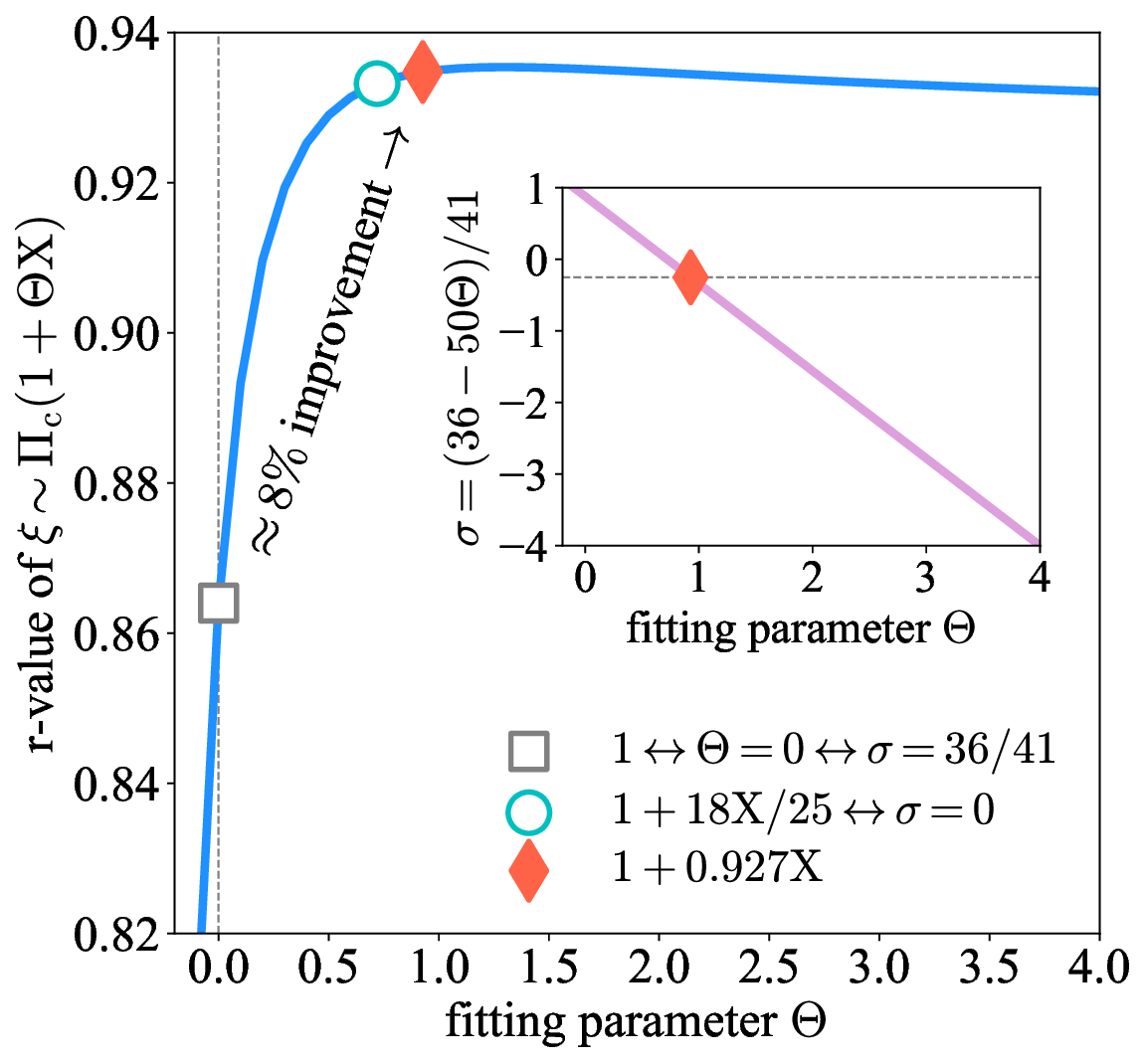}
\caption{(Color Online). The $\Theta$-dependence of the regression $r$-value for the scaling $\xi \sim \Pi_{\rm c}(1+\Theta\x)$, where $\Pi_{\rm c} = \x/(1+3\x^2+4\x)$. 
The inset shows $\sigma = (36 - 50\Theta)/41$ as a function of $\Theta$.
}\label{fig_r_value_fitting}
\end{figure}

In order to assess the improvement in the central EOS-parameter due to the $\sigma$-correction, we show in FIG.\,\ref{fig_rp_xi} the scaling relation between the compactness $\xi_{\max}\equiv M_{\rm{NS}}^{\max}/R_{\max}$ for NSs at the maximum-mass configuration and the term
\begin{equation}\label{ew-ff}
    \frac{\x}{1+3\x^2+4\x}\cdot\frac{1+\Theta\x}{1+2^{-1}\Theta},
\end{equation}
where the fitting parameter $\Theta$ is introduced and the constant $1+2^{-1}\Theta$ is included to make the plots on similar scales.
Since $\Theta\x$ is treated as a perturbation, it is implicitly assumed to satisfy $\Theta\x \lesssim 1$.
A total of 284 realistic EOSs are included in this analysis, broadly classified into: (a) nucleonic models (both microscopic and phenomenological); (b) hybrid EOSs with hyperons and/or $\Delta$ resonances; and (c) quark matter EOSs\,\cite{CLZ23-a}. Particularly, we include here: (1) EOSs with a first-order phase transition, such as APR\,\cite{Akmal1998}, CMF-based EOSs (DS-CMF series)\,\cite{Dexheimer2021PRC}, and VQCD EOSs\,\cite{Jokela2021PRD}; (2) EOSs with a continuous hadron-quark crossover, e.g., AFL series\,\cite{Alford2008} and QHC EOSs\,\cite{Baym19}; and (3) EOSs exhibiting multiple peaks in the $s^2$, as realized in the quark-meson-coupling (QMC) model\,\cite{Stone2021MNRAS,Leong2024NPA}, RMF models with hyperons\,\cite{Stone2021MNRAS}, or sequential QCD phase transitions\,\cite{Alford2017PRL}. For further details on these EOSs, see Refs.\,\cite{Typel2015,Ofeng24}.
It is found from the figure that for $\Theta\gtrsim0.6$, the overall nonlinearity at large $\x$ (see panels (a)-(c)) is largely removed, while negative values of $\Theta$ do not improve the fit. FIG.\,\ref{fig_r_value_fitting} shows the dependence of the fitting r-value on $\Theta$: the optimal fitting occurs around $\Theta\approx1$, where the r-value reaches a maximum of $\approx0.935$ at $\Theta\approx0.927$, compared with the r-value $\approx0.933$ for $\Theta\approx0.72$ (corresponding to $\sigma=0$ or Eq.\,(\ref{ew-0}) for $s_{\rm c}^2$). In contrast, the $\Theta=0$ case gives an r-value of only $\approx0.864$. The overall improvement on the r-value due to the $\sigma$-correction (either $\sigma=0$ or $\sigma\approx-0.253$) compared with the $\Theta=0$ case is about 8\%.
The inset shows the relation between $\sigma$ and $\Theta$, $\sigma=(36-50\Theta)/41$.

Adopting $\Theta\approx1$, we obtain an empirical scaling $\xi_{\max}\sim\x/(1+3\x)$. Numerically, this gives
\begin{equation}
\xi_{\max}\approx\frac{2\alpha}{3}\frac{\x}{1+3\x}+\beta,~~\alpha\approx1.54,~~\beta\approx0.09.
\end{equation}
Since $2\alpha/3\approx1$ and $\beta\approx0.1$, we can write a simple empirical formula for the compactness of NSs at maximum-mass configurations:
\begin{equation}
\text{empirical relation: }
\xi_{\max}\lesssim\frac{\x}{1+3\x}+0.1.
\end{equation}
Using $\x\lesssim0.385$, e.g., then yields $\xi_{\max}\lesssim0.276$, which is consistent with previous studies\,\cite{CL25-a}.
While the $\Theta$ in (\ref{ew-ff}) is an effective fitting parameter, the correction factor $(36-41\sigma)/50\approx0.927\x$  (under $\sigma\approx-0.253$) has a physical origin.

After obtaining the upper bound on $\x$, we can equivalently express the corresponding lower bound on the dimensionless trace anomaly\,\cite{Fuji22}; that is
\begin{equation}
\boxed{
\phi = P/\varepsilon \lesssim 0.385 \leftrightarrow  \Delta \equiv 1/3 - \phi \gtrsim -0.051.}
\end{equation}
This lower bound on $\Delta$ is in good agreement with existing constraints, see, e.g., the plot summarized in Ref.\,\cite{CL25-a}.

\section{Alternative Correction Forms}\label{SEC_AT}

In the previous section, we adopted the effective correction in the form $f(\widehat{r})=-\sigma \x B\widehat{r}^2$ in the expansion of $\hP$. 
Physically, different forms of higher-order corrections are expected to produce qualitatively similar effects on the leading-order scaling relations. In this section, we provide a more detailed analysis of this issue.
It should be emphasized that the correction terms introduced below are not intended as a systematic higher-order perturbative expansion of the dimensionless TOV equations. Instead, they serve as effective parameterizations designed to assess the sensitivity of the scaling relations to plausible deviations beyond the leading-order treatment. Their validity is therefore evaluated primarily through a consistency check with full TOV solutions rather than through an order-by-order perturbative convergence test.

First, we consider the following correction form:
\begin{equation}
f(\widehat{r})=-\varphi\left(B\widehat{r}^2\right)^2 \sim \widehat{r}^4,
\end{equation}
where $\varphi$ is an effective parameter;
then, since $B=6^{-1}(1+3\x^2+4\x)<1$ and $\widehat{R}\lesssim\mathcal{O}(1)$, the magnitude of $(B\widehat{r}^2)^2$ is also expected to be small.  
In this case, the dimensionless radius from the pressure equation $P(R)=\x-B\widehat{R}^2-\varphi B^2\widehat{R}^4=0$ is
\begin{equation}
\widehat{R}^2=\frac{\sqrt{1+4\x\varphi}-1}{2B\varphi}\approx\frac{\x}{B}\left(1-\varphi\x\right).
\end{equation}

The coefficient $a_2$ remains unchanged, $a_2=b_2/s_{\rm{c}}^2=-B/s_{\rm{c}}^2$, since this correction enters at order $\widehat{r}^4$.  
The NS mass then scales as
\begin{align}
M_{\rm{NS}}\sim&\frac{1}{\sqrt{\varepsilon_{\rm{c}}}}\left(\frac{\sqrt{1+4\x\varphi}-1}{2B\varphi}\right)^{3/2}
\cdot\left(1-\frac{3}{5}\frac{B}{s_{\rm{c}}^2}\frac{\sqrt{1+4\x\varphi}-1}{2B\varphi}\right),
\end{align}
while the $s_{\rm c}^2$ remains given by Eq.\,(\ref{ew-3}) as we concern only the linear term in $\x$ in the correction.
Using $\d M_{\rm{NS}}/\d\varepsilon_{\rm{c}}=0$, the perturbative expansion of $s_{\rm{c}}^2$ over $\x$ gives
\begin{equation}
s_{\rm{c}}^2\approx\frac{4}{3}\x+\left(-\frac{2\kappa}{11}+\frac{5\varphi}{33}+\frac{88}{75}\right)\x^2,
\end{equation}
and comparison with Eq.\,(\ref{ew-4}) determines $
\kappa={\varphi}/{10}$.
Consequently, the $s_{\rm c}^2$ becomes (keeping only the linear term in $\x$ in the final correction):
\begin{align}
s_{\rm{c}}^2\approx&\x\left(1+\frac{1}{3}\frac{1+3\x^2+4\x}{1-3\x^2}\right)\left(1+\frac{5\varphi-6}{50}\x\right).
\end{align}

The coefficient $A(\x)=B/s_{\rm{c}}^2$ and the parameter $\varphi$ can be determined by requiring the value of $\x_+$ from $s_{\rm{c}}^2=1$ and the $\overline{\x}$ from $\d^2A(\x)/\d\x^2=0$ with $A(\x)=B(\x)/s_{\rm{c}}^2$ to coincide, which gives $
\varphi\approx0.96$, $\x\lesssim0.375$, as well as $\Delta\gtrsim-0.042$.
Then, the NS mass and radius scale as
\begin{align}
R\sim&\frac{\Pi_{\rm c}^{1/2}}{\sqrt{\varepsilon_{\rm{c}}}}\cdot\left(1-\frac{\varphi}{2}\x\right),~~
M_{\rm{NS}}\sim\frac{\Pi_{\rm c}^{3/2}}{\sqrt{\varepsilon_{\rm{c}}}}\cdot
\left(1+\frac{18-15\varphi}{25}\x\right),
\end{align}
so that
\begin{equation}
\xi\sim\Pi_{\rm c}\cdot\left(1+\frac{36-5\varphi}{50}\x\right),
\end{equation}
with the numerical correction about $0.624\x$.
The scaling and r-value are expected to be similarly good as shown previously, see FIG.\,\ref{fig_r_value_fitting}.

\renewcommand*\tablename{\small TAB.}
\begin{table}[h!]
\centerline{\normalsize
\begin{tabular}{c||c|c|c|c|c|c|c|c} 
\hline
$\sigma$&$-0.4$&$-0.3$&$-0.2$&$-0.1$&0.0&0.1&0.2&0.3\\\hline\hline
$\varphi$&$-0.58$&$-0.16$&0.23&0.61&0.96&1.28&1.61&1.92\\\hline
$\overline{\x}\approx\x_{+}$&0.391&0.386&0.382&0.379&0.375&0.372&0.369&0.366\\\hline
$\Theta$&1.11&0.98&0.86&0.74&0.62&0.51&0.39&0.28\\\hline
\end{tabular}}\caption{Some representative values. A range $0.7\lesssim\Theta\lesssim1.0$ is favored for the fitting, here $\Theta=18/25-41\sigma/50-\varphi/10$; for a given $\sigma$, the $\varphi$ is determined by requiring $\overline{\x}\approx\x_+$.}\label{tab_aa} 
\end{table}

For a general correction of the form $f(\widehat{r})\sim(B\widehat{r}^2)^n$ with $n\geq3$, since $B\widehat{R}^2\lesssim1$, the effect decreases with increasing $n$.  
Thus the linear form $-\sigma\x B\widehat{r}^2$ provides the largest possible contribution to relevant quantitie.
Including both corrections, $-\sigma\x B\widehat{r}^2$ and $-\varphi(B\widehat{R}^2)^2$, gives $\x-(1+\sigma\x)B\widehat{R}^2-\varphi(B\widehat{R}^2)^2=0$, so
\begin{equation}
\widehat{R}=\frac{1}{\sqrt{2\varphi B}}\left[\sqrt{(1+\sigma\x)^2+4\varphi\x}-(1+\sigma\x)\right]^{1/2},
\end{equation}
with $
\kappa=11\sigma/50+\varphi/10$ determined in a similar manner, and
\begin{align}
R\sim&\frac{\Pi_{\rm c}^{1/2}}{\sqrt{\varepsilon_{\rm{c}}}}\cdot\left(1-\frac{\sigma+\varphi}{2}\x\right),\\
M_{\rm{NS}}\sim&\frac{\Pi_{\rm c}^{3/2}}{\sqrt{\varepsilon_{\rm{c}}}}\cdot
\left(1+\frac{18-33\sigma-15\varphi}{25}\x\right),\\
\xi\sim&\Pi_{\rm c}\cdot\left(1+\frac{36-41\sigma-5\varphi}{50}\x\right).
\end{align}
See TAB.\,\ref{tab_aa} for selected values for $\varphi$ determined by requiring $\overline{\x}\approx\x_+$ and the parameter $\Theta=18/25-41\sigma/50-\varphi/10$ (for the compactness scaling), when $\sigma$ changes from $-0.4$ to $0.3$. We notice that $0.7 \lesssim \Theta \lesssim 1.1$ is required to ensure a high r-value for the fitting as shown by  FIG.\,\ref{fig_r_value_fitting}, and in this region the upper bound for $\x$ is roughly in the range of 0.39 to 0.37. This shows that the upper bound $\x \lesssim 0.385$ obtained in the previous section, using only the $\sigma$-correction, is quite reasonable. We emphasize that the coincidence between the causality and mass-sphere bounds is imposed here as a self-consistency condition within the IPAD-TOV framework, rather than as an independent prediction. Accordingly, the value $\x \lesssim 0.385$ should be regarded as the preferred upper bound obtained with the adopted effective parameterization, while the spread of values in TAB.\,\ref{tab_aa} provides an estimate of the associated systematic uncertainty.

Finally, for a general correction of the form
\begin{equation}
f(\hr)=-\sigma\x B\hr^2+\sum_{j=1}\varphi_j(B\hr^2)^{2j},
\end{equation}
the reduced radius becomes
\begin{align}
&\widehat{R}^2\approx\frac{\x}{B}\frac{1}{1+\sigma\x}\cdot
\left(1-\frac{\sum_j\varphi_j\x^{2j}}{\x+2\sum_j j\varphi_j\x^{2j}}\right),
\end{align}
so that the radius scaling becomes
\begin{align}
R\sim&\frac{\Pi_{\rm c}^{1/2}}{\sqrt{\varepsilon_{\rm{c}}}}\cdot\left(1-\frac{\sigma\x}{2}-\frac{1}{2}\frac{\sum_j\varphi_j\x^{2j}}{\x+2\sum_j j\varphi_j\x^{2j}}\right)\notag\\
\approx&\frac{\Pi_{\rm c}^{1/2}}{\sqrt{\varepsilon_{\rm{c}}}}\cdot\left(1-\frac{\sigma\x}{2}-\frac{\varphi_1\x}{2}\right).
\end{align}
This demonstrates that the leading-order correction in $\x$ remains unchanged, and that higher-order terms (characterized by $\varphi_j$ with $j\geq2$) do not affect the scaling of the linear term in $\x$ within the correction.

Overall, the present analysis suggests that the leading-order scaling relations are insensitive to the detailed form of higher-order corrections. Nevertheless, the effective parameterizations adopted here should be viewed as phenomenological rather than physically unique. They do not constitute a complete systematic expansion beyond quadratic order, and their applicability is ultimately limited to the range over which they remain consistent with full numerical TOV solutions.

\section{Summary and Conclusions}\label{SEC_SUM}

The physical information encapsulated in the coefficient $A(\x)\equiv -a_2(\x)>0$ with $\x=P_{\rm c}/\varepsilon_{\rm c}$, in the perturbative expansion of the reduced energy density $\heps\approx1-A(\x)\hr^2+\cdots$ in the IPAD-TOV approach is revealed through a Gedankenexperiment. Physically, $A(\x)$ decreases with $\x$ at small $\x$ with a positive second-order derivative $\d^2A/\d\x^2$; the critical value $\overline{\x}$ defined by $\d^2A/\d\x^2=0$ signals the onset of mass-sphere instability: further increasing $\x$ beyond $\overline{\x}$ accelerates the rise of the mass-sphere near the center, which indicates an unstable state. The value $\overline{\x}\approx 0.377$ is close to the causality-only upper bound $\x_+$ for $\x$ about 0.374, demonstrating the self-consistency and partially explaining the effectiveness of IPAD-TOV scalings for NS mass, radius, and compactness, even when truncated at low orders.

Building on this insight, we derive an improved upper limit on the central EOS-parameter $\x$ within the IPAD-TOV framework by incorporating the mass-sphere stability condition near the NS center. This refinement modifies the expression for $s_{\rm c}^2$, yielding $\x\lesssim0.385$. Although this upper limit is obtained within the present perturbative treatment rather than as a general proof, it remains consistent with the causality-only estimate, improves the NS compactness scaling across 284 realistic microscopic EOSs, and is further supported by existing microscopic EOS calculations and astrophysical observations.
Consequently, the dimensionless trace anomaly at NS centers, $\Delta_{\rm c}=1/3-\x$, is bounded from below by $\gtrsim -0.05$. In the present work, the effective correction to $s_{\rm c}^2$ is kept at the linear order $1+\Theta\x$ with an optimal $\Theta\approx0.927$. Future studies may aim to work out the higher-order form, $1+\Theta\x+\Theta_2\x^2+\cdots$, where terms such as $\Theta_2\x^2$ are expected to be small since $\x\lesssim 0.38$-$0.39$, indicating that the upper bound on $\x$ obtained in this work remains robust and practically useful.
 
The upper limit on $\x$ provides a useful benchmark for the maximum compression achievable in NS cores within the present IPAD-TOV framework. Although it is not a fundamental consequence of causality alone, its consistency with microscopic EOS calculations, full TOV solutions, and current astrophysical observations suggests that it captures an empirically robust feature of dense matter under strong gravitational fields. Further refining this limit may therefore improve our understanding of the properties of superdense matter at the highest densities realized in nature.

\section*{Acknowledgment} We would like to thank Rui Wang and Zhen Zhang for helpful discussions. This work was supported in part by the National Natural Science Foundation of China under contract No. 12547102, the U.S. Department of Energy, Office of Science, under Award Number DE-SC0013702, the CUSTIPEN (China-U.S. Theory Institute for Physics with
Exotic Nuclei) under the US Department of Energy Grant No. DE-SC0009971.

\section*{Data Availability}
The data that support the findings of this article will be openly available \cite{data}.

\end{document}